\documentclass[aps,prl,twocolumn,showpacs,amsmath,amssymb,superscriptaddress]{revtex4}

\usepackage{graphicx}
\usepackage{epstopdf}
\usepackage{amsmath}
\usepackage{amssymb}
\usepackage{booktabs}
\usepackage{color}
\usepackage{multirow}

\begin{document}

\title{The interplay of sedimentation and crystallization in hard-sphere suspensions} 

\author{John Russo}
\email{russoj@iis.u-tokyo.ac.jp}
\affiliation{ {Institute of Industrial Science, University of Tokyo, 4-6-1 Komaba, Meguro-ku, Tokyo 153-8505, Japan} }
\author{Anthony C. Maggs}
\email{anthony.maggs@espci.fr}
\affiliation{ {PCT, ESPCI, 10 Rue Vauquelin, 75005 Paris, France} }
\author{Daniel Bonn}
\email{D.Bonn@uva.nl}
\affiliation{ {van der Waals-Zeeman Institute, University of Amsterdam, Valckenierstraat 65, 1018 XE Amsterdam, The Netherlands} }
\author{Hajime Tanaka}
\email{tanaka@iis.u-tokyo.ac.jp}
\affiliation{ {Institute of Industrial Science, University of Tokyo, 4-6-1 Komaba, Meguro-ku, Tokyo 153-8505, Japan} }

\begin{abstract}
We study crystal nucleation under the influence of sedimentation in a model
of colloidal hard spheres via Brownian Dynamics simulations. We introduce two
external fields acting on the colloidal fluid: a uniform gravitational field (body force),
and a surface field imposed by pinning a layer of equilibrium particles (rough wall).
We show that crystal nucleation is suppressed in proximity of the wall due to the slowing
down of the dynamics, and that the spatial range of this effect is governed by the static length scale
of bond orientational order. For distances from the wall larger than this length scale,
the nucleation rate is greatly enhanced by the process of sedimentation, since it leads to a higher volume fraction, or a higher degree 
of supercooling, near the bottom. The nucleation stage
is similar to the homogeneous case, with nuclei being on average spherical and having crystalline
planes randomly oriented in space. The growth stage is instead greatly affected by the
symmetry breaking introduced by the gravitation field, with a slowing down
of the attachment rate due to density gradients, which in turn cause nuclei to grow faster laterally. 
Our findings suggest that the increase of crystal nucleation in higher density regions might be the cause of the large discrepancy 
in the crystal nucleation rate of hard spheres between experiments and simulations, on noting that the gravitational effects 
in previous experiments are not negligible.
\end{abstract}

\maketitle

\section{Introduction}
Crystal nucleation is a fundamental physical process whose understanding has far-reaching
consequences in many technological and industrial products, like pharmaceuticals, enzymes
and foods~\cite{kelton2010,palberg,anderson,AuerR,SearR,GasserR,karayiannis2012spontaneous}. The simplest crystallization process is the homogeneous nucleation case,
in which solid clusters spontaneously form from the melt throughout the system. The opposite
case is instead the heterogeneous nucleation process, where nuclei of the solid phase form preferentially
around external surfaces, like containers walls or impurities present in the melt~\cite{cacciuto2004onset,winter2009monte}.
But the crystallization processes in practical systems are often very far from these idealized cases,
for example when multiple fields concurrently affect the crystallization behaviour, making it difficult to
match theoretical expectations with experimental outcomes. Quoting the famous words of
Oxtoby~\cite{oxtoby}, ``nucleation theory is one of the few areas of science in which agreement
of predicted and measured rates to within several orders of magnitude is considered a major success''.
The most idiomatic example comes from the simplest crystallization process,
the homogeneous crystallization of hard spheres, where the discrepancy between predicted nucleation
rates and experimental measurements stretches as far as 10 orders of magnitude.
In particular, numerical simulations using a variety of
techniques (Brownian dynamics, biased Monte Carlo, and rare-events methods) found that the nucleation
rate increases dramatically with the colloid volume fraction $\phi$, growing by more than 15 orders of magnitude from
$\phi=0.52$ to $\phi=0.56$, where it has a maximum~\cite{auer2001prediction,zaccarelli,filion,kawasaki,pusey2009hard,filion2,schilling_jpcm,valeriani2012compact}. On the other side, experiments found the nucleation rate 
to be much less sensitive on the volume fraction~\cite{schatzel1993density,harland1997crystallization,sinn2001solidification,iacopini,franke2011heterogeneous}.
This is probably the second worst prediction in physics, the first being the $100$ orders of
magnitude difference between the cosmological constant predicted from the energy of the
vacuum and that measured from astronomical data~\cite{hobson2006general}.

In the present work we address a very important factor affecting the crystallization process, which often occurs in real experiments of
colloidal suspensions but has been ignored in most simulations: how the crystallization process is affected by the sedimentation of particles.
We perform Brownian Dynamics simulations of a model of colloidal hard spheres, and induce sedimentation by introducing
both a gravitational force $G$ and rough walls which confine the system along the direction of gravity.
The effects of rough walls on both the static and dynamic properties of the colloidal fluid are analyzed in detail.
In particular we will show that there is strong slowing down of the dynamics close to the walls, and that this effect
has a static origin. Correspondingly, the crystallization process is strongly suppressed in proximity of the walls, which
allow us to study the nucleation process under gravity without interference from the walls. We will show in fact that both
the nucleation rates, crystal shape and the orientation of crystalline planes are similar to what observed at bulk
conditions. On the other hand, the gravitation field strongly affects the growth stage, and we will show that
nuclei grow more slowly across a density gradient, and thus prefer to grow laterally.
This indicates that not only local density but also its gradient affect the crystallization behavior.

We also provide new insights on the debated origin of the discrepancy between theoretical predictions and
experimental measurements of nucleation rates in hard spheres.
We first note that the experiments measuring the nucleation rates in hard spheres
are usually characterized by rather short gravitational lengths
(and quite marked sedimentation effects have indeed been
reported~\cite{schatzel1993density,underwood1994sterically}). We will then present some arguments
to show that sedimentation should have a rather big effect in these experiments, especially at lower volume fractions, 
where the discrepancy is much more significant.

The paper is organized as follows. In Section~\ref{sec:methods} we describe the methods employed in
our study and the choice of the state points considered. Section~\ref{sec:results} presents
the results of the study, logically divided in five parts.
Section~\ref{sec:nucleation_rates}
examines the effects of gravity on the nucleation rates measured by simulation.
Section~\ref{sec:nucleation} deals with
the effects of gravity on the static properties of the suspension. Section~\ref{sec:wall_effects} investigates
the effects of the walls, both on the dynamics and the statics. Section~\ref{sec:growth} considers instead the growth of the nuclei as affected by gravity.
Section~\ref{sec:experiments} compares our results with previous experimental investigation of the
crystallization in hard-sphere colloidal systems.
We conclude in Section~\ref{sec:conclusions}.

\section{Methods}\label{sec:methods}
We perform Brownian Dynamics (BD) simulations of spherical particles interacting through the
 Weeks-Andersen-Chandler (WCA) potential \cite{weeks1971}

\[
\beta U(r) = \left\{
  \begin{array}{l l}
    4\beta\epsilon\left( \left( \frac{\sigma}{r} \right)^{12} -\left( \frac{\sigma}{r} \right)^6 +\frac{1}{4}\right) & \quad \text{for $\frac{r}{\sigma}\leq 2^{1/6}$}\\
    0 & \quad \text{for $\frac{r}{\sigma}>2^{1/6}$}\\
  \end{array} \right.
\]

\noindent
where $\sigma$ is the length scale, $\epsilon$ is the energy scale and $\beta=1/k_{\rm B}T$ ($k_{\rm B}T$: the thermal energy).
In the following we set the energy scale to $\epsilon=1$.
The WCA potential is a purely repulsive short-range potential. The value of $\beta$ fixes the hardness of the interaction, and we
choose $\beta=40$ for which a mapping to the hard-sphere phase diagram is known. In particular in Ref.~\cite{filion2} the
freezing density was located at $\rho_F=0.712$, which can be compared to the volume fraction
of hard spheres at the freezing transition ($\phi_F=0.492$) to define an effective hard-sphere volume ($v_\text{eff}$) for WCA particles,
$\rho_F\,v_\text{eff} = \phi_F$. The mapping of the WCA system onto the HS phase diagram
is then simply given by the relation
\begin{equation}
 \rho_\text{WCA}\,v_\text{eff}=\phi_\text{HS}. 
\end{equation}
The effective hard-sphere diameter $d$ of our particles is then given by $d=\sqrt[3]{6\phi_\text{HS}/\pi\rho_\text{WCA}}\sim 1.097\sigma$.

In BD the equation of motion of particle $i$ is
$$
\frac{d\mathbf{r}_i}{dt}=\frac{D}{k_{\rm B}T} \mathbf{f}_i + \boldsymbol{\eta}_i(t),
$$
where $t$ is the time, $\mathbf{r}_i$ is the position of particle $i$, $D$ is the bare diffusion coefficient, $\mathbf{f}_i$ is the systematic force acting on particle $i$ and $\boldsymbol{\eta}_i$
is the noise term describing the effective stochastic force exerted by the solvent on particle $i$ and obeying the fluctuation-dissipation  relation
$\langle\boldsymbol{\eta}_i(t)\boldsymbol{\eta}_j(t')\rangle=6D\delta_{ij}\delta(t-t')$. In the following we set $D/k_{\rm B}T=1$ and
integrate the equations of motion by the standard Ermak integrator~\cite{allen} with a time step of $\Delta t=10^{-5}\sigma^2/D$. The Brownian
time $\tau_B=d^2/D$ is the time it takes for a colloid to diffuse a distance equal to its diameter in a dilute suspension.

The systematic force acting on particle $i$ has two terms
$$
\mathbf{f}_i=-\boldsymbol\nabla_iU + \mathbf{f}_B\,
$$
where the first term accounts for the conservative forces between the particles, and the second term is the body force,
given by the difference between the gravitational force and the buoyancy force
$$
\mathbf{f}_B=v_\text{eff}\left(\rho_f-\rho_P\right)\,\hat{\mathbf{z}}\equiv -G\,\hat{\mathbf{z}},
$$
where $\rho_f$ is the density of the implicit solvent into which the particles are suspended,
$\rho_P$ is the density of the colloidal particles, $G$ is the modulus of the total
body force, and $\hat{\mathbf{z}}$ is the unit vector opposite to the direction of gravity.

The gravitational force, breaking the translational symmetry in the $z$ direction, produces a $z$-dependent
density profile, also called barometric law $\rho(z)$~\cite{piazza1993equilibrium}.
This density profile can be calculated from the pressure difference between two altitudes $z_i$ and $z_j$ as
\begin{equation}\label{eqn:stevino}
 p(z_i)-p(z_j)=-G\int_{z_j}^{z_i}\rho(z)\,dz,
\end{equation}
by inserting the appropriate equation of state, $p(\rho)$, on the left hand side. We use the Carnahan-Starling equation of state
$$
\beta\,p = \frac{\rho\left(1+\phi+\phi^2-\phi^3\right)}{(1-\phi)^3},
$$
where $\phi=\rho\,v_\text{eff}$ is the volume fraction. Equation~(\ref{eqn:stevino}) can then be rewritten as
an integral equation whose solution is given in an implicit form by the roots of the following equation
\begin{equation}\label{eqn:profile_1}
 \log\phi+\frac{1}{(\phi-1)^2}-\frac{2}{(\phi-1)^3}=-\beta G z+K,
\end{equation}
where $K$ is a constant fixed by the following normalization condition
\begin{equation}\label{eqn:profile_2}
 \int_0^h\phi(z;K)\,dz=h\,\phi_\text{avg},
\end{equation}
where $h$ is the height of the simulation box (in the direction of the gravitational field) and $\phi_\text{avg}$ is the
volume fraction averaged over the total volume occupied by the particles in the simulation box.
The theoretical determination of the density profile inside the simulation box thus requires fixing the field $G$, the height $h$,
the average volume fraction $\phi_\text{avg}$ occupied by the particles in the simulation box, and solving numerically Eq.~(\ref{eqn:profile_1}) and
Eq.~(\ref{eqn:profile_2}).

\begin{figure}[!t]
 \centering
 \includegraphics[width=6.5cm,clip]{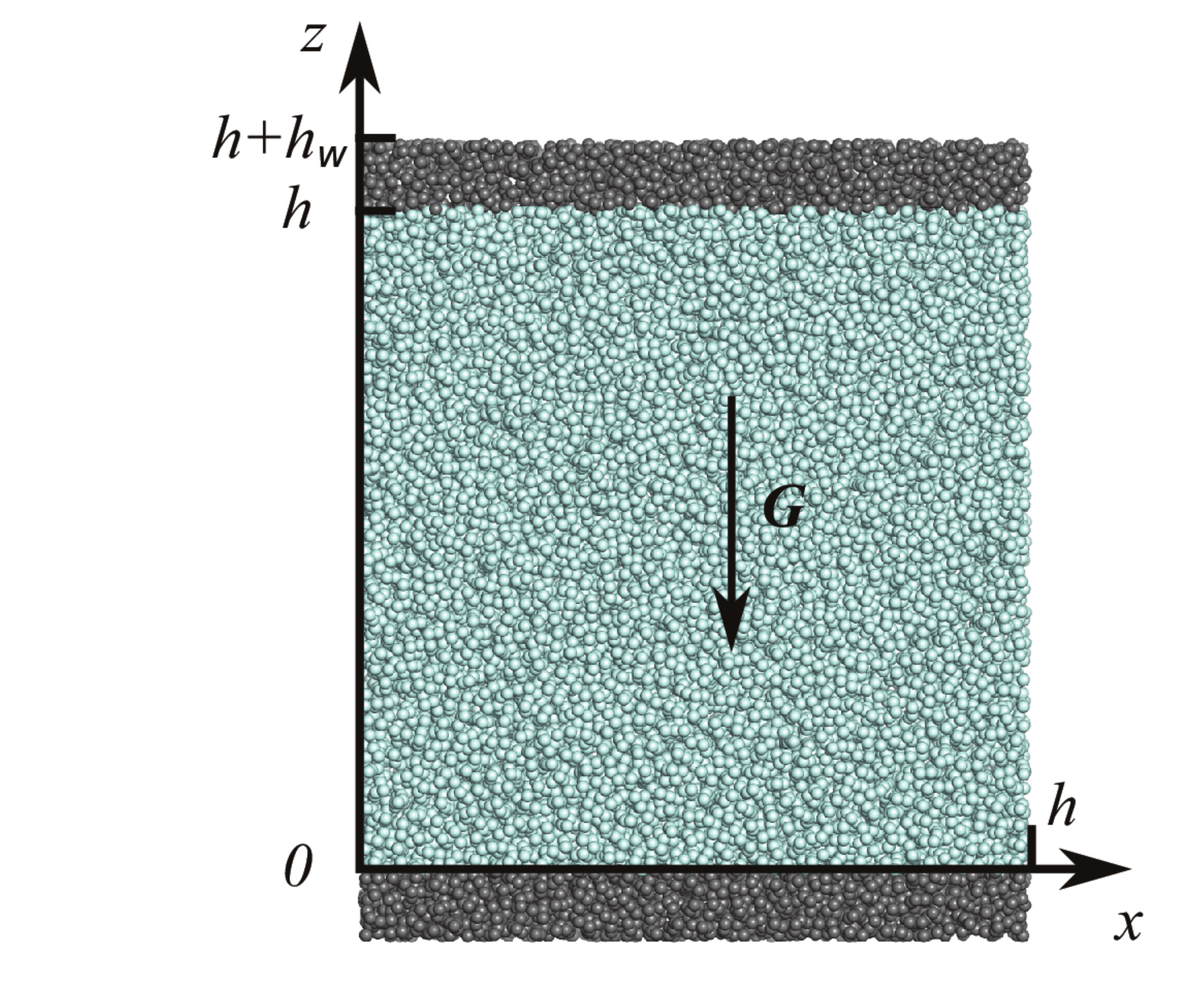}
 \caption{Simulation box configuration. The fluid is confined between $z=0$ and $z=h$ by two rough walls of height
$h_W$ each. The external force (with body acceleration of $G$) acts in the negative $\hat{\mathbf{z}}$ direction.
Wall particles are depicted by dark (gray) spheres, while fluid particles are depicted by light (blue) spheres.
We choose $h_W=3\sigma$ and the box length along $x$ (and $y$) equal to $h$.}
 \label{fig:box}
\end{figure}

Before applying the external force we need to bound the system with walls in the direction perpendicular to the external field.
Choosing flat walls would induce heterogeneous nucleation, whereas we want to study the homogeneous process which
happens in the bulk in the presence of the external field. We then choose to confine our systems with rough walls, obtained by freezing
the positions of particles in equilibrated fluid configurations.
We will show in Section~\ref{sec:wall_effects}, that rough walls indeed disfavour nucleation in their
proximity and are thus the appropriate choice for our investigation. It is also well known that rough walls do not induce layering effects, as the fluid's density
remains unperturbed in their proximity~\cite{scheidler2002cooperative,scheidler2004relaxation}. Ideally we wish thus to prepare the walls at the same state 
point of the layer of fluid in contact with the wall. Since the external field will induce a density gradient in the system we thus need
to predict the density of the fluid at $z=0$ and $z=h$ ($h$ being the height of the box, see Fig.~\ref{fig:box}).

A representation of the simulation box is depicted in Fig.~\ref{fig:box}. The protocol for the simulations is as follows.
\begin{description}
 \item[profile prediction]- given $G$, $h$ and $\phi_\text{avg}$ we solve Eqs.~(\ref{eqn:profile_1}) and (\ref{eqn:profile_2}) to obtain the density profile $\rho(z)$.
 \item[wall preparation]- two independent BD simulations are run respectively at $\rho(0)$ and $\rho(h)$ in the absence of the external field. Since the predicted
$\rho(0)$ is often very high, nucleation could occur at this stage, so we add a biasing potential $U_\text{bias}$ which prevents the systems from
nucleating. The biasing potential has the form of $U_\text{bias}=k n^2$, where $k$ is an harmonic constant and $n$ is the size of the largest crystal
in the box at each time step. 
 \item[box setup]- a slab of height $h_W$ is cut from each of the two previous configurations. The slab at density $\rho(0)$ is placed at $-h_W<z<0$
of the new simulation box, whereas the slab at density $\rho(h)$ is placed between $h<z<h+h_W$. $N$ fluid particles are placed randomly 
between $0<z<h$ at the 
volume fraction $\phi_\text{avg}$, and then equilibrated with the external field switched off. A typical simulation box is depicted in Fig.~\ref{fig:box}.
 \item[simulation run]- at $t=0$ the field $G$ is switched on and simulations are run until the size of the largest nucleus reaches $n_\text{max}=500$.
The position of wall particles is kept fixed.
\end{description}

We set $N=20,000$ fluid particles (not
including wall particles), with the height $h$ equal to the box dimensions in both the $x$ and $y$ directions, for which periodic boundary
conditions are imposed.

\subsection{Identification of crystal particles}
To identify crystal particles we use the local bond-order analysis introduced by
Steinhardt et al.~\cite{steinhardt}, first applied to study crystal nucleation by
Frenkel and co-workers~\cite{auer}. 
A $(2l+1)$ dimensional complex vector ($\mathbf{q}_l$)
is defined for each particle $i$ as $q_{lm}(i)=\frac{1}{N_b(i)}\sum_{j=1}^{N_b(i)} Y_{lm}(\mathbf{\hat{r}_{ij}})$, where
$l$ is a free integer parameter, and $m$ is an integer
that runs from $m=-l$ to $m=l$. The functions $Y_{lm}$ are the spherical harmonics
and $\mathbf{\hat{r}_{ij}}$ is the vector from particle $i$ to particle $j$.
The sum goes over all neighbouring particles $N_b(i)$ of particle $i$. Usually 
$N_b(i)$ is defined by all particles within a cutoff distance, but in an inhomogeneous system
the cutoff distance would have to change according to the local density. Instead we 
fix $N_b(i)=12$ which is the number of nearest neighbours in close packed crystals (like hcp and fcc)
which are known to be the only relevant structures for hard spheres.
If the scalar product $(\mathbf{q}_6(i)/|\mathbf{q}_6(i)|)\cdot(\mathbf{q}_6(j)/|\mathbf{q}_6(j)|)$ between
two neighbours exceeds $0.7$ then the two particles are deemed \emph{connected}. We then identify particle $i$ as crystalline
if it is connected with at least $7$ neighbours.
A useful order parameter which is built from the previous bond-order analysis is
\begin{equation}\label{eqn:crystallinity}
 \text{S}_i=\sum_{j=0}^{N_b(i)}\frac{\mathbf{q}_6(i)\cdot\mathbf{q}_6(j)}{|\mathbf{q}_6(i)|\,|\mathbf{q}_6(j)|}.
\end{equation}
It measures the coarse-grained bond orientational order of particle $i$, which is a very effective order parameter to
measure the coherence of crystal-like bond orientational order. Hereafter we call this ``crystallinity''~\cite{russo_hs}. 
However, we note that it is not a direct indicator for the presence of crystals, but rather a measure 
for a tendency to promote crystallization.

\subsection{Gravitational length and time scales}
The gravitational field breaks the translational symmetry of the system and introduces a characteristic
length scale called the gravitational length, $l_G$. The gravitational length describes the typical length scale over
which the density profile decreases in the $z$ direction. For a dilute gas the density profile is given by the barometric law
$\phi(z)\sim e^{-G z /k_{\rm B}T}$, and thus
\begin{equation}
 l_G=\frac{k_{\rm B} T}{G}
\end{equation}
where $G$ is the effective gravitational force. To compare to the experiments, we report the adimensional length $l_G/d$ (see also below), where
$d$ is the hard-sphere diameter of the particles.

In addition to the length scale, the gravitational field defines also a time scale, the \emph{sedimentation time} $\tau_S$,
which is the time it takes for a particle to move over the distance $d$ due to the gravitational pull.
The velocity attained by a sphere pulled by the gravity inside a fluid is simply given by $v_\text{drag}=G/\zeta$, where
$\zeta$ is the drag coefficient (which can be computed from the viscosity of the solvent by using the Stokes law).
The sedimentation time is then given by $\tau_S=d\zeta/G$. The \emph{P\'{e}clet number}, $\text{Pe}$, is given by
the ratio of the diffusion time to the sedimentation time. 
The Brownian time is simply $\tau_B=d^2/D=d^2\zeta/k_{\rm B}T$, and so
\begin{equation}
 \text{Pe}=\frac{\tau_B}{\tau_S}=\frac{d G}{k_{\rm B}T}=\frac{d}{l_G}.
\end{equation}
In our simulations $l_G > d$ and so we are working in the regime of small P\'{e}clet numbers, which is
the relevant regime for colloidal dispersions used in estimating the nucleation rate (see Table \ref{tab:experiments}).
All results reported
in the following sections are taken after waiting for at least $3\tau_S$ before acquiring data.

\subsection{Choice of state points}

\begin{table}[!b]
\small
  \caption{Simulated state points. Each state point is uniquely defined by the definition
of the gravitational length, $l_G$, and the average volume fraction of particles in the simulation
box, $\phi_\text{avg}$. Simulations are divided into four groups.
In Group I all simulations have
approximately the same density at $z=0$ but differ for their gravitational lengths. In Group II simulations
have the same gravitational length but differ for their densities at $z=0$. In Group III the highest density
is still low enough to avoid crystallization during the simulation time. In Group IV all simulations have
the same gravitational length, comparable to some colloidal experiments~\cite{schatzel1993density,sinn2001solidification}.
For simulations in Group I, II and IV
we report the effective nucleation rate, $k\,d^5/D$, and the average height where nucleation occurs,
$\langle z\rangle$.}
  \label{tab:state_points}
  \begin{tabular*}{0.48\textwidth}{@{\extracolsep{\fill}}lllll}
   \hline
\emph{group} & $l_G/d$ & $\phi_\text{avg}$ & $k\,d^5/D$ & $\langle z\rangle/d$  \\
\hline
\multirow{3}{*}{I} & 2.07 & 0.530  & $9.5\cdot 10^{-6}$ & 5.6 \\
 & 1.90 & 0.525 & $6.5\cdot 10^{-6}$ & 5.2 \\
 & 1.75 & 0.520 & $5.7\cdot 10^{-6}$ & 4.9 \\ \hline
\multirow{3}{*}{II} & 1.75 & 0.540  & $1.7\cdot 10^{-5}$ & 7.2 \\
 & 1.75 & 0.520 & $5.7\cdot 10^{-6}$ & 4.9 \\
 & 1.75 & 0.510 & $2.9\cdot 10^{-6}$ & 4.0 \\ \hline
\multirow{3}{*}{III} & 7.59 & 0.530  &  &  \\
 & 5.70 & 0.525 &  &  \\
 & 4.56 & 0.520 &  &  \\ \hline
\multirow{6}{*}{IV} & 3.10 & 0.520  & $7.4\cdot 10^{-7}$ & 3.8  \\
   & 3.10 & 0.525 & $1.0\cdot 10^{-6}$ & 4.2 \\
   & 3.10 & 0.530 & $1.9\cdot 10^{-6}$ & 4.6 \\
   & 3.10 & 0.540 & $4.4\cdot 10^{-6}$ & 6.7 \\
   & 3.10 & 0.550 & $6.1\cdot 10^{-6}$ & 8.8 \\
   & 3.10 & 0.560 & $7.5\cdot 10^{-6}$ & 11.1 \\ \hline

\hline
  \end{tabular*}
\end{table}
The state points simulated in the present work are reported in Table~\ref{tab:state_points}
(the volume comprised by the walls along $z$ and the periodic boundaries along $x$ and $y$ directions 
is cubic, and the height $h$ can be readily obtained from $\phi_\text{avg}$).
The points are divided into the following four groups. 

\begin{description}
\item[\ \ I]: once the profile is settled, these simulations have the same average density at $z=0$ but
differ for their gravitational lengths $l_G$. With these simulations we investigate the effect of the 
strength of the density gradient produced by the gravitational field on the crystallization process.
\item[\ II]: these simulations all have the same gravitational length $l_G$ but differ for their average densities.
With these simulations we can investigate the effect of the walls on the nucleation process.
\item[III]: all simulations have a density low enough to avoid the crystallization of the system, and are thus suited to study the
effect of the gravitational field and of the walls on the dynamics of the melt (or, supercooled liquid) prior to crystallization.
With these simulations we can investigate the effect of the walls on the nucleation process.
\item[\ IV]: these simulations have a gravitational length $l_G$ comparable with that of several experiments in index matched but not
density matched solvents~\cite{schatzel1993density,sinn2001solidification}. This group is used to study the effects of gravity
on the nucleation rates.

\end{description}

\begin{figure}[!b]
 \centering
 \includegraphics[width=8cm,clip]{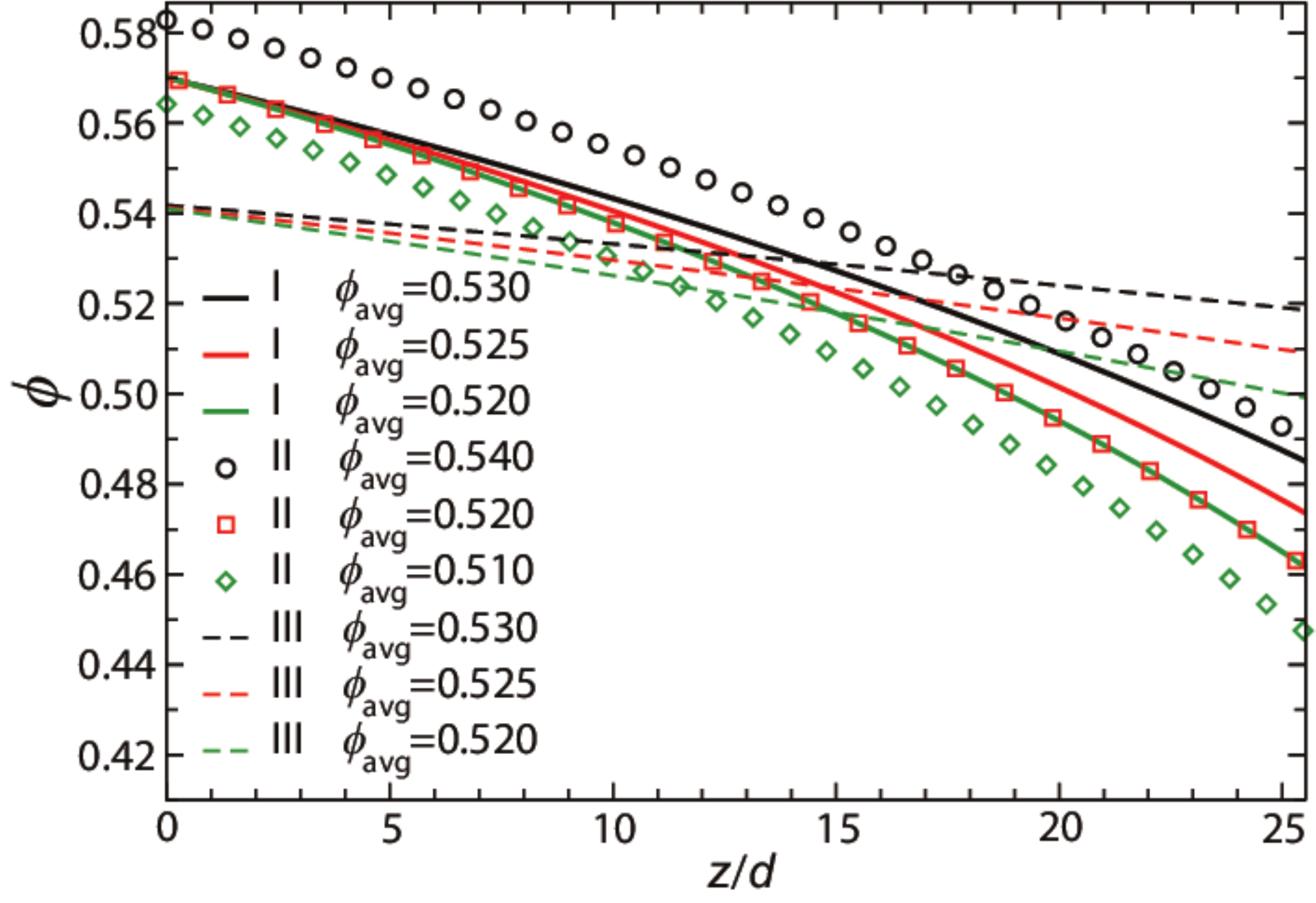}
 \caption{Theoretical volume fraction profiles calculated from Eqs.~(\ref{eqn:profile_1}) and (\ref{eqn:profile_2}) for the state points
of groups I, II and III, in Table~\ref{tab:state_points}. Group I state points are depicted with continuous lines: they are characterized
by $\phi(z=0)\sim 0.570$ and different density gradients. Group II simulations are depicted with open symbols: they all have the
same gravitational length and accordingly the density profiles are parallel. Group III simulations are depicted with dashed lines:
their volume fraction $\phi<0.54$ and thus nucleation events are never observed during our observation time.}
 \label{fig:theo_profiles}
\end{figure}


\begin{figure*}[!t]
 \centering
 \includegraphics[width=13cm,clip]{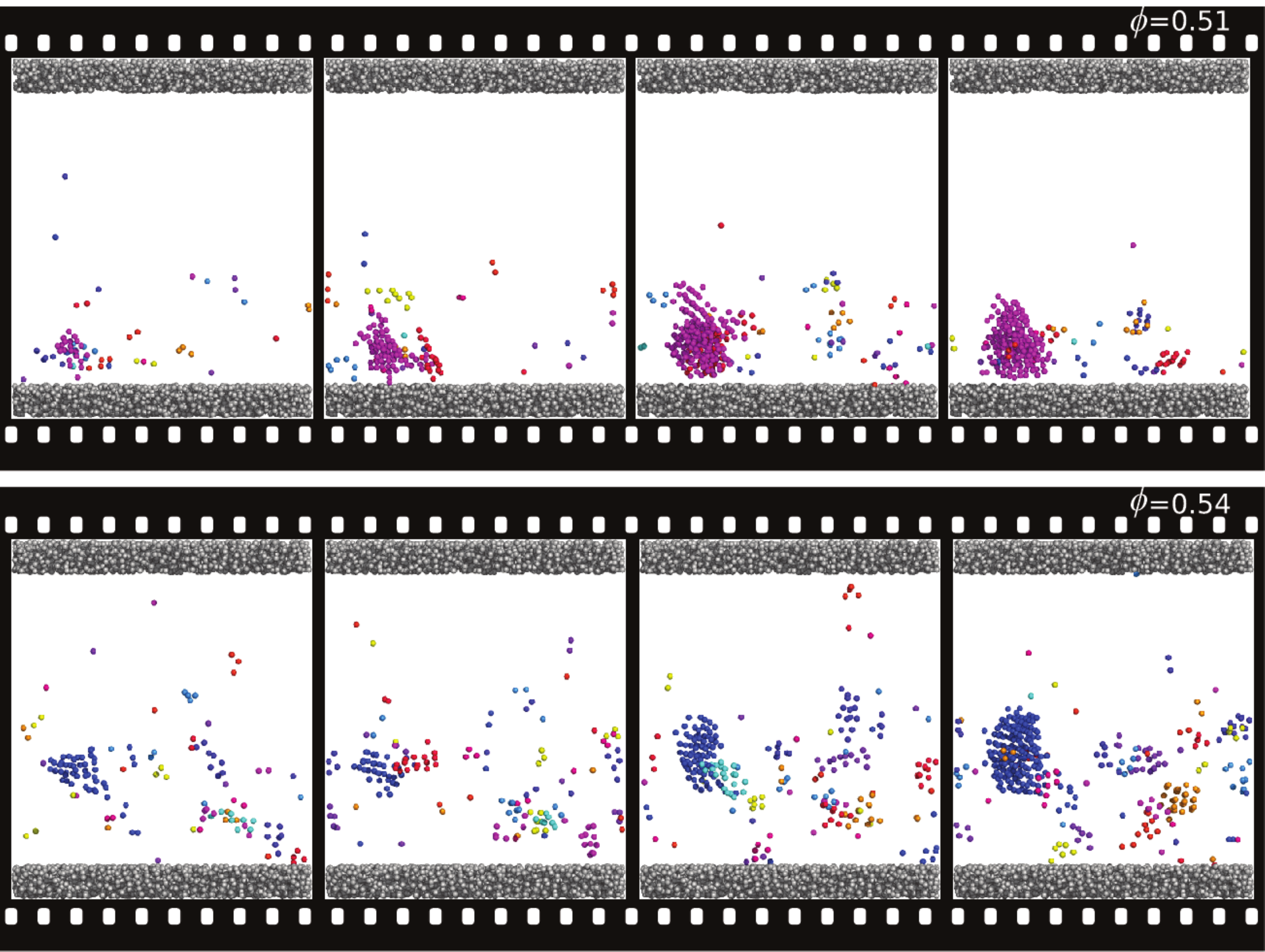}
 \caption{Nucleation snapshots for simulations of group II, at $\phi_\text{avg}=0.510$ (top row) and $\phi_\text{avg}=0.540$ (bottom row). Wall particles are coloured in grey, 
whereas only crystalline particles are shown and coloured according to the cluster they belong to. An algorithm is used to identify particles belonging
to the same cluster in time, so that the colouring of the clusters should remain consistent across the time frames. For $\phi_\text{avg}=0.510$ snapshots are 
taken at a time interval of $\Delta t=3\tau_B$ after waiting for $t_0=3\,\tau_S$ to ensure the settling of the profile. For $\phi_\text{avg}=0.510$ snapshots are taken at a time interval of $\Delta t=1.5\tau_B$ after waiting for $t_0=3\,\tau_S$.
The snapshots span clusters from pre-critical to post-critical sizes.}
 \label{fig:snapshots}
\end{figure*}

We show the theoretical profiles calculated from Eqs.~(\ref{eqn:profile_1}) and (\ref{eqn:profile_2}) for the state points
in group I, II, III in Fig~\ref{fig:theo_profiles}.
State points with the same gravitational length are characterized by the same density gradient across the box. Decreasing the
gravitational length increases the density gradient.

\section{Results}\label{sec:results}

\subsection{Nucleation rates}\label{sec:nucleation_rates}
We directly measure nucleation rates in our simulations by running $50$ independent simulations for each state point in groups I, II and IV (see Table~\ref{tab:state_points}).
In the absence of a gravitation field, the nucleation rate as calculated by simulations has a very strong dependence on the density,
growing by 15 orders of magnitude by just going from $\phi=0.52$ to $\phi=0.54$~\cite{AuerN} (see Fig. \ref{fig:nucleationrate}). 
The direct simulation of
nucleation events becomes unfeasible for $\phi<0.53$ and one has to resort to rare-events sampling techniques in order to extract
the nucleation rate~\cite{AuerN,filion}. This is not the case in the
presence of a gravitational field: most of our simulation state points are within $\phi_\text{avg}<0.53$ but still we are able to observe directly
nucleation events, for all state points of groups I, II and IV. For the calculation of the nucleation rate $k$ we resort to the direct formula
\begin{equation}
 k=\frac{1}{\langle t\rangle V},
\end{equation}
where $\langle t\rangle$ is the average time at which nucleation events occur, and $V$ is the system's volume. The nucleation rate
of course depends sensibly on the definition of the nucleation time. We define the nucleation time as the time it takes for the
largest nucleus in the system to reach size $100$ particles. This size is bigger than the critical nucleus size, as all nuclei that reach
this size always keep growing. For the volume $V$ we use the volume available to the fluid, even if (as we will see later) nucleation
events do not occur in the whole volume. Despite the fact that both the choice of the critical size and of $V$ are very conservative, potentially
leading to lower nucleation rates than actually observed, the nucleation rates reported in Table~\ref{tab:state_points} are
very high, comparable to the nucleation rates which homogeneous systems have around the nucleation rate maximum, at $\phi\sim 0.56$.
A great enhancement of the nucleation rates is indeed observed in our systems. In the following section we
will address the origin of this enhancement, and whether the nucleation stage is really akin to a homogeneous nucleation process.

\subsection{Static properties}\label{sec:nucleation}
Previous studies have addressed the crystallization of hard spheres in gravity by confining the system with flat walls~\cite{volkov2002molecular,hoogenboom2003real,mori2006crystal,marechal2007crystallization,ramsteiner2009experimental,allahyarov2011huge}. In this case the high nucleation
rates were due to heterogeneous nucleation on the walls. In order to prevent heterogeneous nucleation, we
confine our system with rough walls, i.e. walls that are obtained by freezing a zone of colloidal particles, occupying positions that are 
characteristic of the bulk liquid. It is well known that such frozen walls do not induce the density layering typical of flat smooth  walls~\cite{scheidler2002cooperative,watanabe_walls}. This is due to the fact that the roughness leads to the lack of the phase coherence of 
the density waves.  

As a first step to prove that walls are not enhancing our nucleation rate, we run simulations of the WCA fluid confined by rough walls prepared at volume
fraction of $\phi_w=0.5657$ in the absence of gravity. The fluid within the walls was prepared at different volume fractions, from $\phi=0.54$ to
$\phi=0.57$, and nucleation events were seen to occur randomly in the simulation box, without any apparent enhancement in the 
proximity of the walls.

When a gravitational field is turned on, a density profile is induced in the simulation box. We first start by visually locating the nucleation
events, as shown in Fig.~\ref{fig:snapshots}. From these direct observations we can already infer that the location of the nucleation events
depends sensibly on the local density. For $\phi_\text{avg}=0.510$ (top row) nucleation occurs very close (but not in contact) with the wall, while
for $\phi_\text{avg}=0.540$ (bottom row) it is located too far away to
be due to wall effects. While always distinct, many nucleation events can occur in the simulation box, a consequence
of the high nucleation rate, and in principle interactions between the different nuclei will occur.

\begin{figure}[!t]
 \centering
 \includegraphics[width=8cm,clip]{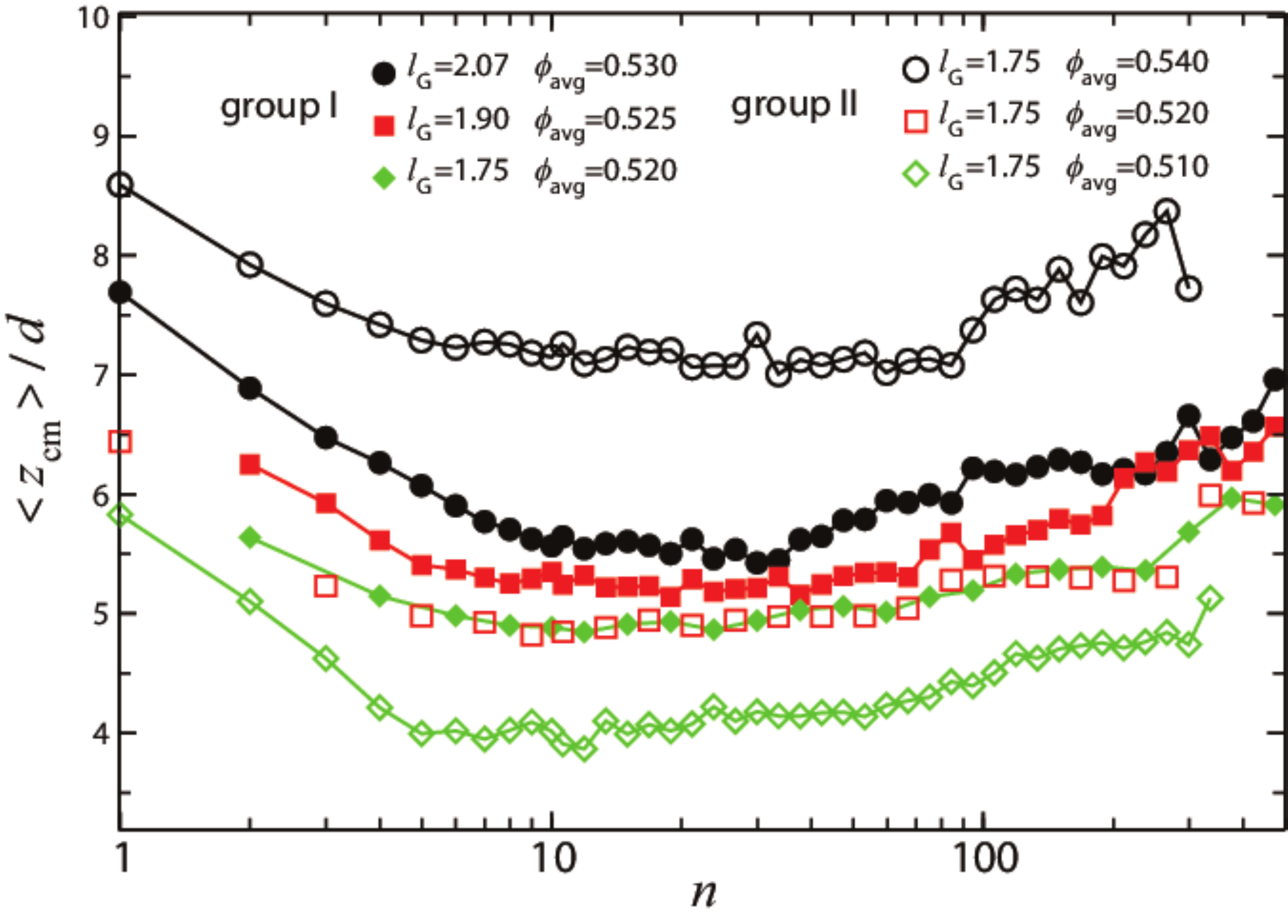}
 \caption{Average height of the centers of mass of nuclei as a function of their size, for all state points of group I (closed symbols) and
group II (open symbols). Averages are done separately for each nucleus size, and then sizes within the same histogram bin (in logarithmic scale) are averaged together. The average height displays a clear plateau at intermediate sizes, which corresponds to the average height $\langle z\rangle$ of nucleation events and is reported in Table~\ref{tab:state_points}.}
 \label{fig:avz}
\end{figure}

To study these events in detail we determine the average location of the nucleation events, $\langle z\rangle$, which is summarized also in Table~\ref{tab:state_points}.
To calculate $\langle z\rangle$ we first detect all individual nuclei via a cluster algorithm, and then calculate the average height of the centers of mass as
a function of the size of the nucleus $n$. The results are reported in Fig.~\ref{fig:avz}. For each state point, the average height of the centers of mass has a characteristic
dependence on the size of the nucleus. For very small nuclei ($n\lesssim 10$) the height of the center of mass decreases with $n$: this is due to the fact that small
nuclei randomly form in a large portion of the simulation box, so that their average height is high, while growing nuclei form
preferentially at the bottom of the simulation box. With increasing $n$, $\langle z\rangle$ reaches a plateau which encompasses
the critical nucleus size and can thus be considered as the average height at which nucleation events occur. The average nucleation height is clearly correlated
with the density profile of each state point, as we will see shortly.
Interestingly, for $n\gtrsim 60$ the average height increases
again, which means that the growth of nuclei occurs on average more in the positive $z$ direction, thus opposite to the direction of 
gravity.

\begin{figure}[!t]
 \centering
 \includegraphics[width=8cm,clip]{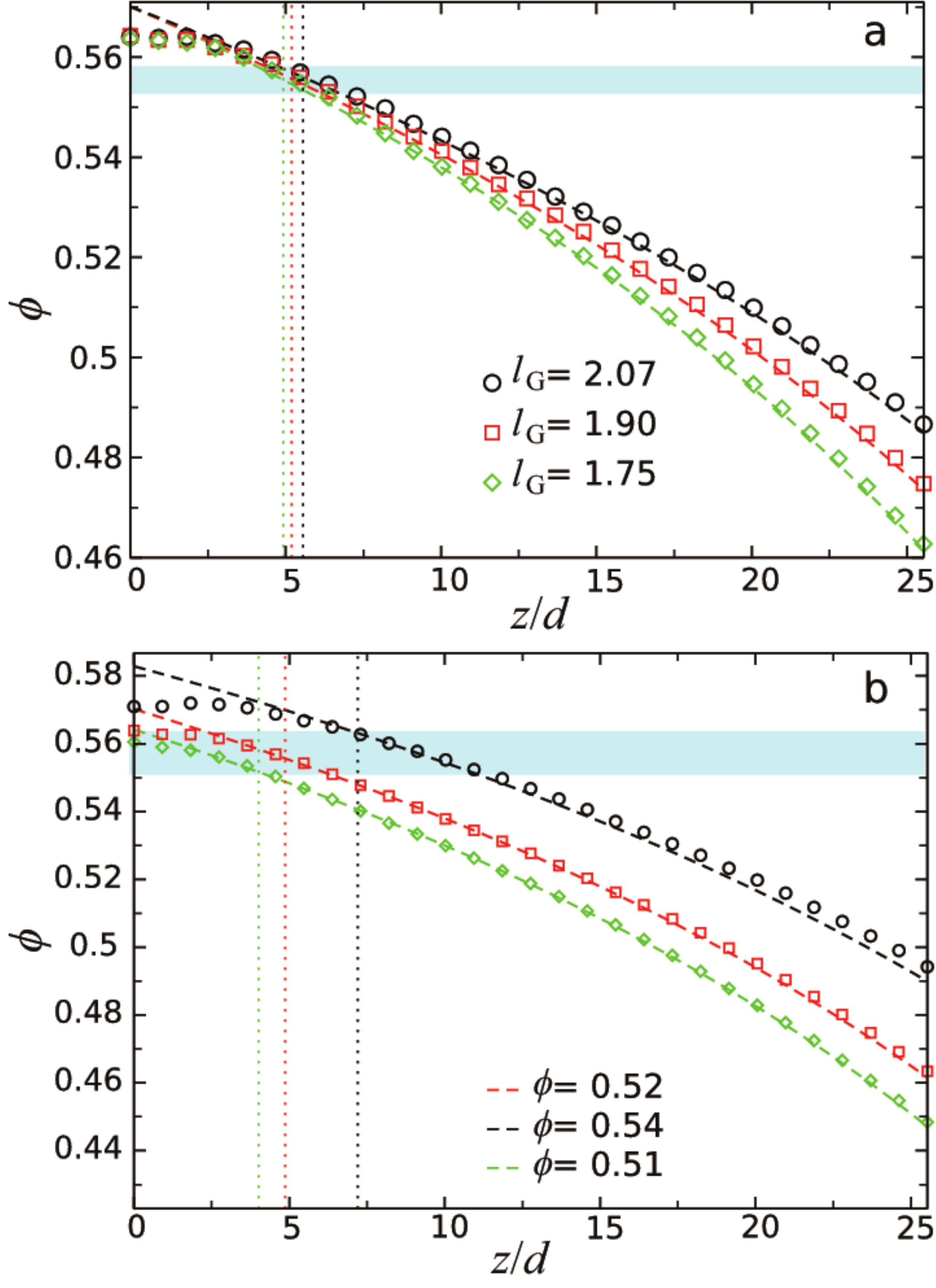}
 \caption{Volume fraction profiles $\phi(z)$ for group I (a) and group II (b) simulations, obtained by means of Voronoi diagrams. The simulated profiles are represented by symbols, while
dashed lines are theoretical predictions based on Eqs.~(\ref{eqn:profile_1}) and (\ref{eqn:profile_2}). The vertical dotted lines show the average height of nucleation as determined
from the plateaus in Fig.~\ref{fig:avz}. The coloured horizontal band in both figures represents the
$\phi$ region where average nucleation events occur, as determined by the intersection of the density profiles with the vertical dotted lines.
Simulation profiles are calculated by averaging configurations with the biggest nucleus having
size between $50$ and $60$ particles, and by dividing the $z$ dimension into bins of size $\Delta z=1$.}
 \label{fig:phi_profile}
\end{figure}

Figure~\ref{fig:phi_profile} plots the volume fraction profile $\phi(z)$ for group I (a-top panel) and group II (b-bottom panel) state points.
The measured profile (symbols) is obtained by averaging over configurations where the biggest nucleus is of size between $50$ and $60$ particles,
thus capturing the profile just before the growth stage.
The measured profile (symbols) can be compared with the expected equilibrium profiles, calculated from Eqs.~(\ref{eqn:profile_1}) and (\ref{eqn:profile_2}),
and plotted in Fig,~\ref{fig:phi_profile} as dashed lines. For all state points we note that the actual profile at the time of nucleation
is in very good agreement with the equilibrium one for distances not too close to the wall ($z=0$). Next to the walls, instead
the density saturates to a constant value. Density profiles are practically unchanged also even at later times, when nuclei have started filling the system.
In Fig.~\ref{fig:phi_profile} we also report as dotted vertical lines the average height of nucleation events, as determined in Fig.~\ref{fig:avz}.
By intersecting these lines with the corresponding density profiles we note that all nucleation events (irrespective of $\phi_\text{avg}$ and gravitational length)
occur in regions where $0.55\lesssim\phi(z)\lesssim 0.56$. This interval is exactly the volume fraction 
where the nucleation rate in bulk has a maximum.
It is thus clear that the origin of the high nucleation rates, and the localization of the nucleation events, corresponds
to homogeneous nucleation occurring in regions characterized by a local volume fraction of $0.55\lesssim\phi(z)\lesssim 0.56$.
In the next section we will provide an explanation for the saturation of the density profile close to the walls, but in the meanwhile we 
emphasize that nucleation events occur in regions of the simulation box where density has relaxed.

\begin{figure}[!t]
 \centering
 \includegraphics[width=8cm,clip]{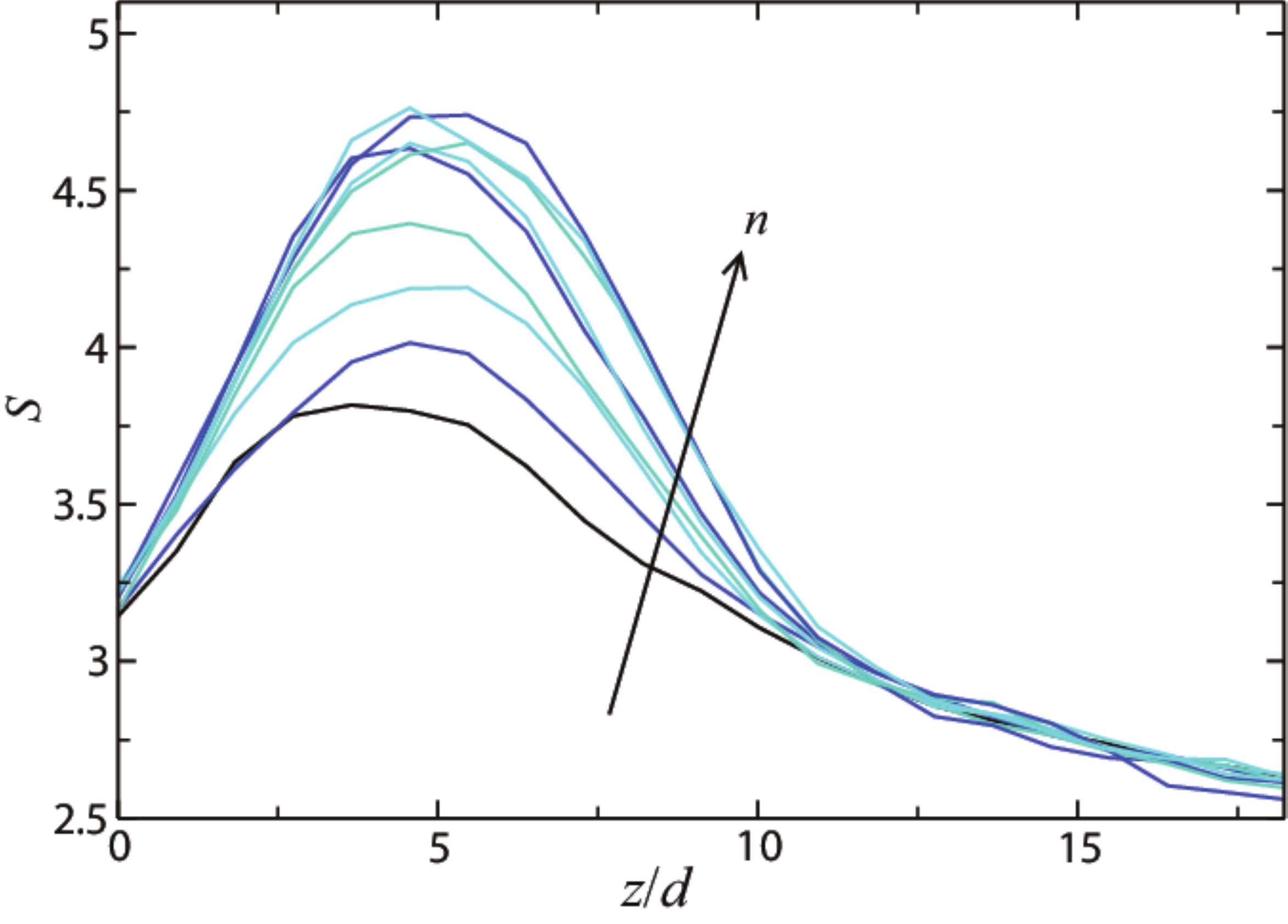}
 \caption{Average crystallinity order parameter, as defined by Eq.~(\ref{eqn:crystallinity}), for the state point of group I and $l_G=1.9$. The different curves 
are averages of the crystallinity for configurations with nuclei of size $n\pm 5$, for the following values of $n=25$, 55, 105, 155, 205, 355, and 405 (the order
is specified by the arrow). The crystallinity profile increases rapidly with the size of the growing nuclei.}
 \label{fig:crystallinity}
\end{figure}

By looking at the density profiles it is difficult to detect the presence of the growing nuclei, since the density change 
between the small nuclei and
the fluid phase is very small. Moreover, growing nuclei are known to have a density closer to the melt than to the 
bulk crystal up to sizes many times
larger than the critical nucleus size~\cite{russo_hs}. Growing crystals are more easily detected by bond orientational order parameters, 
such as the
one introduced in Eq.~(\ref{eqn:crystallinity}) which we refer to as \emph{crystallinity} order parameter~\cite{russo_gcm,russo_hs}. 
A plot of the profile for this
order parameter is shown in Fig.~\ref{fig:crystallinity} for the state point of group I with $l_G=1.9$.
The different curves show the average profile for configurations with embedded nuclei of different
size $n$ (we take $n$ as the size of the largest nucleus in each configuration). As the size of the nucleus $n$ grows, the crystallinity rapidly increases.
The average mean position of the crystalline peak is in good agreement with the one extracted from Fig.~\ref{fig:avz}. Also we note that the average
peak position shifts to higher values of $z$ as the size of the nuclei grow, as was also observed in Fig.~\ref{fig:avz}. We thus once again
confirm that, along the $z$ direction, the growth of the nuclei occurs preferentially opposite to the gravitational force.

We conclude this section by raising two questions. The first one is why the density does not relax to its equilibrium value close to the walls, even long after
nucleation has started. A second question, possibly related to the first one, is why nucleation never occur close to the wall. 
As for this last question, let us take as an example the state point of group II and $\phi_\text{avg}=0.510$. As can be seen from Fig.~\ref{fig:phi_profile}(b) the
nucleation starts on average at a distances around $4\sigma$ from the wall, despite the fact that the density approaches $\phi=0.56$ going closer to the wall, 
where the nucleation rate should have its maximum.
To answer these questions, in the next section we study the effects of rough walls on the static and dynamical properties of the fluid.

\subsection{Wall effects}\label{sec:wall_effects}
The effects of walls on the static and dynamical properties of fluids is of great interest, and many studies have been devoted
to this problem~\cite{scheidler2002cooperative,scheidler2004relaxation,watanabe_walls,kob2012non}.
To study the combined effects of gravity and rough walls we use state points of group III in Table~\ref{tab:state_points}, where nucleation events do not occur within the simulated time.

\begin{figure}[!t]
 \centering
 \includegraphics[width=8cm,clip]{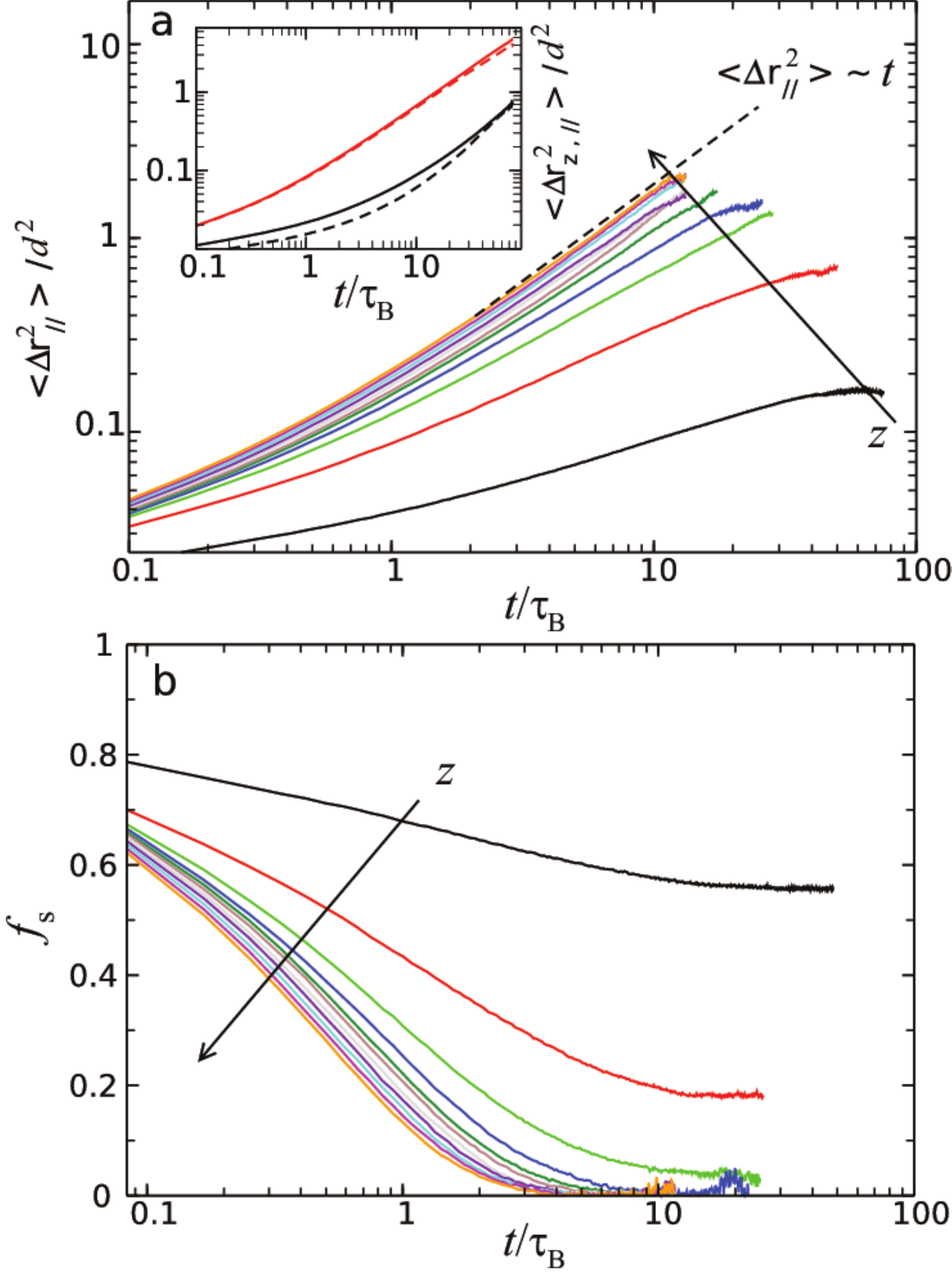}
 \caption{Lateral mean square displacement ((a) - top panel) and intermediate scattering function ((b) - bottom panel) as a function of the
distance $z$ from the wall, for the state point of group III and $\phi_\text{avg}=0.53$. We divide the systems into slabs of $\Delta z=\sigma$ and calculate the mean square displacement (a) and the intermediate scattering function (b) for those trajectories which do not leave 
the slab. The different lines correspond to the following values $z/\sigma=1$, 2, 3, 4, 5, 7, 9, 11, 13, and 15 and the order is 
given by the arrow. The inset in panel (a) shows the lateral mean
square displacement ($\langle\Delta r_{//}^2\rangle/2$ - continuous lines) and the perpendicular mean square displacement
($\langle\Delta r_{z}^2\rangle$ - dashed lines) for those trajectories starting at $z=0$ (black lines) and $z=4\sigma$ (red lines). 
Unlike the main panel (a), these trajectories are allowed to leave the slab. 
}
 \label{fig:dynamics}
\end{figure}

We start by looking at the dynamics. In Fig.~\ref{fig:dynamics}(a) we plot the lateral mean square displacement for trajectories 
belonging to parallel slabs
at distance $z$ from the wall. We compute the lateral mean square displacement according to the following formula~\cite{liu_diffusion}
$$
\langle\Delta r^2_{//}(t)\rangle_z\!=\!\!\frac{1}{N_z(t)}\!{\sum_{z<z_i<z+\sigma}}\!\!\!\!\!\!\!{\left(x_i(t)-x_i(0)\right)^2\!+\!\left(y_i(t)-y_i(0)\right)^2},
$$
where $N_z(t)$ is the number of particles which are in the slab $[z,z+\sigma]$ at time $t$. The figure clearly shows that the lateral motion of the particles is slower as 
we approach the wall. For $z\lesssim 3\sigma$ the mean square displacement does not reach the diffusive regime, $\langle\Delta r^2_{//}(t)\rangle\sim t$, and
for the slab at $z=0$ the motion is still sub-diffusive even after $100$ Brownian times. For $z\gtrsim 4\sigma$ the mean square displacement eventually
reaches the diffusive regime, with a diffusion constant which grows as $z$ increases. The increase of diffusivity as a function of $z$ is clearly due to the
decrease of density with $z$. Since the growth of nuclei is controlled by diffusion, which determines the rate
at which particles attach to the crystalline seed, we can firmly predict that the growth of nuclei will be faster on the side of 
the nucleus far from the
wall. This is exactly what we saw in Fig.~\ref{fig:avz} and Fig.~\ref{fig:crystallinity}, where the centers of mass of the nuclei moves 
toward higher $z$ as
they grow. The inset in Fig.~\ref{fig:dynamics}(a) compares the lateral mean square displacement (continuous lines) to the one in the $z$ direction (dashed lines)
for particles starting their trajectories in slabs at $z=0$ (black lines) and $z=4\sigma$ (red lines). Whereas for the slab at $z=4\sigma$ the mean square displacement
is isotropic in all directions, for the $z=0$ slab the diffusivity in the $z$ direction is lower than the lateral one. Close to the
wall ($z\lesssim 3\sigma$) the diffusion tensor has different components in the $(x,y)$ and $z$ directions, whereas 
isotropy is recovered for $z\gtrsim 4\sigma$.

In Fig.~\ref{fig:dynamics}(b) we plot the intermediate scattering function for density fluctuations in the $(x,y)$ plane
with a wave number $q=|\mathbf{q}|$ corresponding to the first peak in the structure factor, calculated according to 
the following formula: 
$$
f_s(q,t)=\left< \frac{1}{N_z(t)}\!{\sum_{z<z_i<z+1}}e^{-i\mathbf{q}\cdot \binom{x_i(t)-x_i(0)}{y_i(t)-y_i(0)}} \right>. 
$$
The different curves correspond to slabs at different heights. Again, for $z\lesssim 3\sigma$ we can see that the self scattering function
has still not decayed to zero, meaning that density fluctuations are not able to relax in the observed time window.
Near the walls the dynamics slows considerably, and this is at the origin of the non-equilibrium profile observed in the
previous section (Fig.~\ref{fig:phi_profile}) for slabs close to the wall. 

\begin{figure}[!t]
 \centering
 \includegraphics[width=8cm,clip]{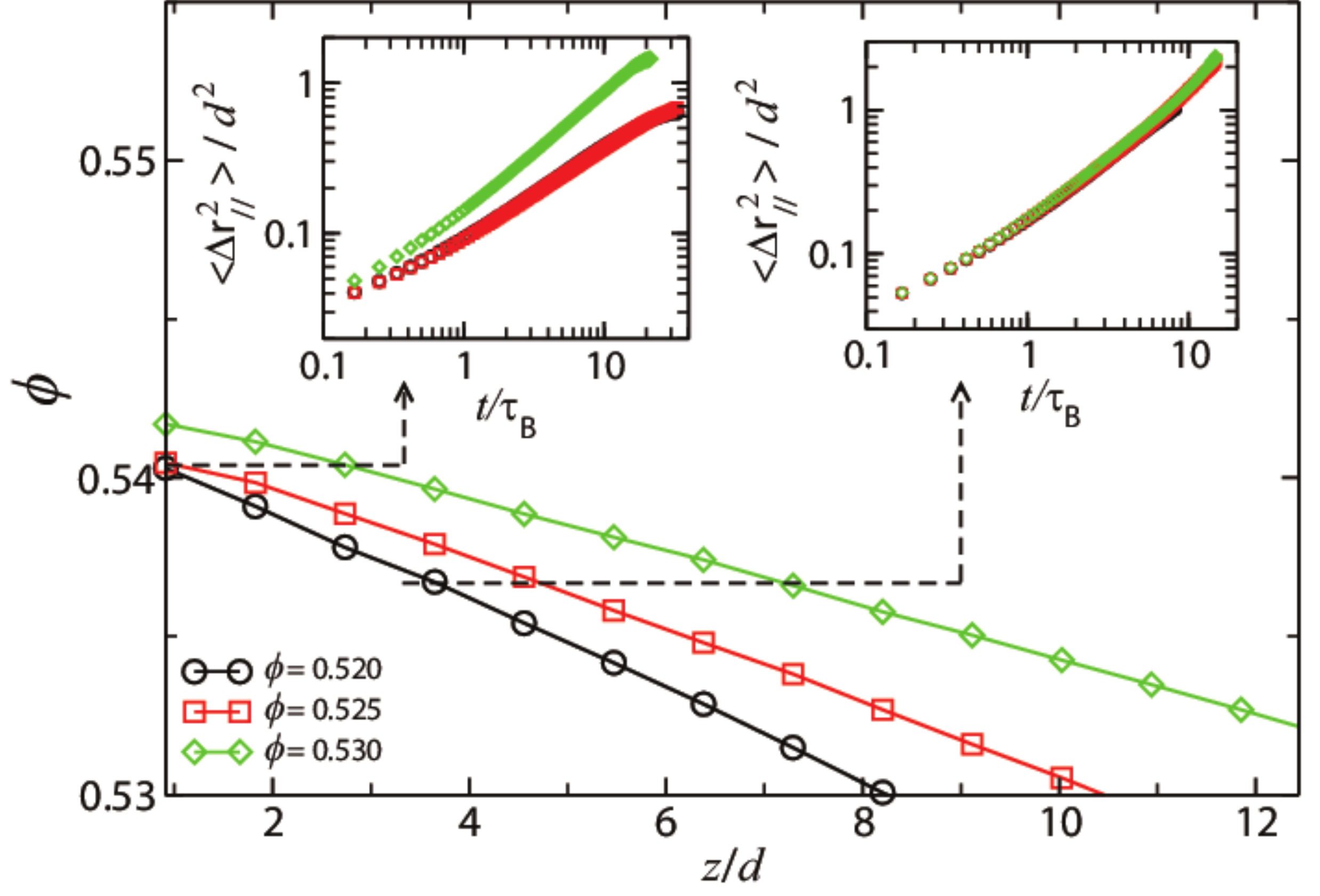}
\caption{Volume fraction profile $\phi(z)$ for state points of group III, and comparison of the lateral mean square displacement for slabs 
at $\phi=0.54$ (left inset) and at $\phi=0.537$ (right inset). Choices of colours and symbols are consistent between the volume fraction profiles in the main panel, and the lateral mean square displacements in the two insets.}
 \label{fig:dynamics_comparison}
\end{figure}

We can investigate the range of the wall effects by comparing the dynamics between simulations with different gravitational lengths.
In Fig.~\ref{fig:dynamics_comparison} we plot the dynamics 
of slabs located at different distances from the wall but with the same local volume fraction. The main 
panel shows the volume fraction profiles $\phi(z)$ for the group III state points. For each of the state points, 
the dynamics of slabs having the average density of $\phi=0.54$ (left inset) and $\phi=0.537$ are then compared. 
In the right inset all slabs are at distance $z\geq 4\sigma$ and they 
display the same dynamics. Thus the dynamics is bulk-like for  $z\gtrsim 4\sigma$, and independent
of the local density gradient. For $\phi=0.54$ (left inset), the dynamics of the slabs located at $z=1\sigma$ is much slower than the dynamics
at $z=3\sigma$. We can again conclude that for $z\lesssim 3\sigma$ there are strong wall effects.

\begin{figure}[!t]
 \centering
 \includegraphics[width=8cm,clip]{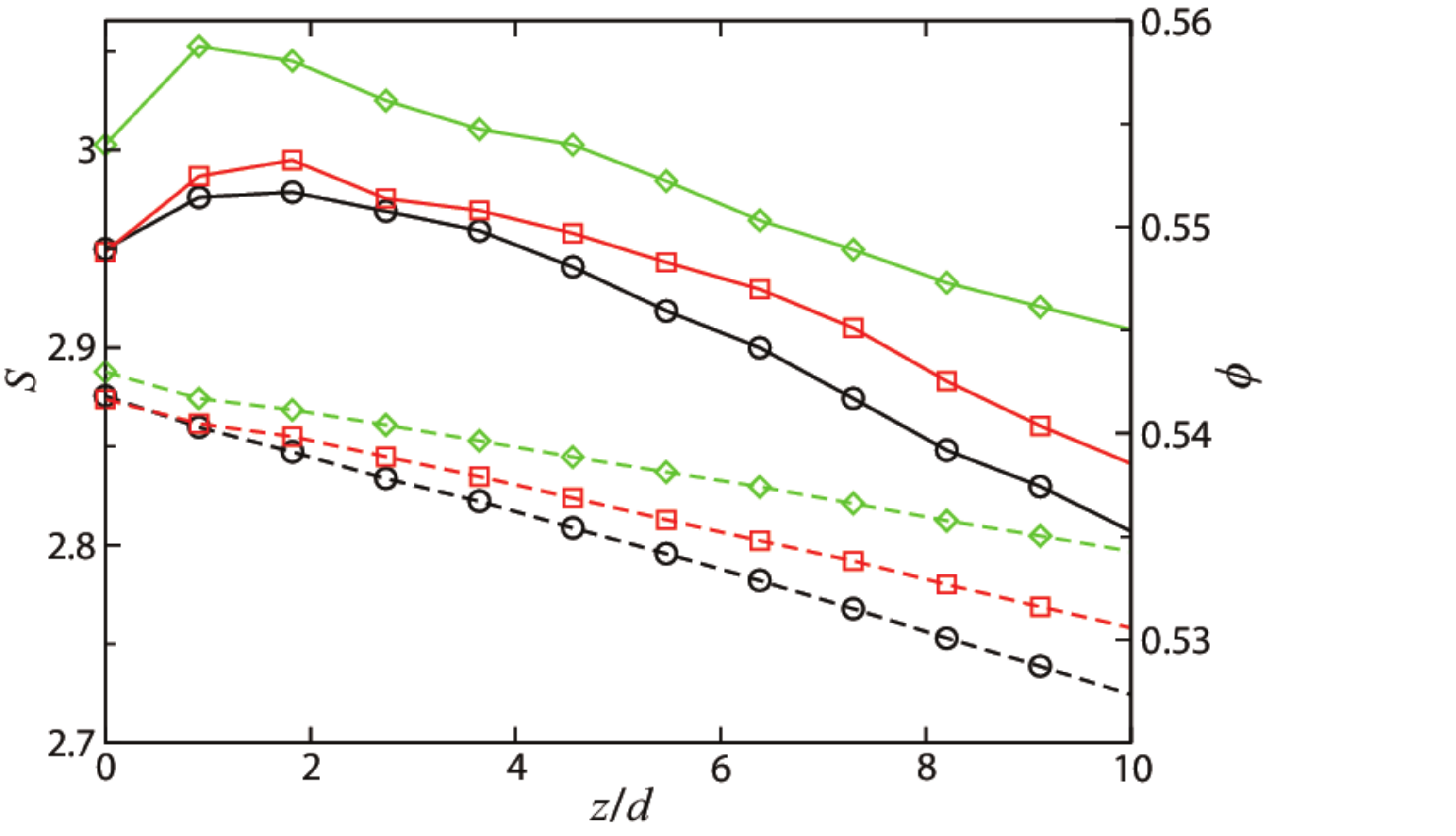}
 \caption{Crystallinity (solid lines, left axis scale) and volume fraction (dashed lines, right axis scale) 
profiles for state points of group III. Results are averaged by taking slabs with $\Delta z=\sigma$.}
 \label{fig:walls_statics}
\end{figure}

We now proceed to study the effects of the walls on the static properties of the fluid. We consider positional order (as expressed by
the local density) and bond orientational order (expressed by the crystallinity order parameter defined in Eq.~(\ref{eqn:crystallinity})),
both depicted in Fig.~\ref{fig:walls_statics}. Both translational and bond orientational order grow as $z$ decreases, but their behavior
in the proximity of wall is very different. While density is almost unperturbed on approaching the wall, crystallinity is instead
strongly suppressed. The range of this suppression coincides well with the region where deviations from bulk dynamics were observed.
A link between static and dynamic properties under confinement was recently proposed in Ref.~\cite{watanabe_walls}, 
and it is compatible with our
findings. Moreover it was recently argued that crystallization is driven by bond orientational order and not 
by positional order~\cite{russo_hs}, 
and this is clearly shown in our results: while density is rather unperturbed on approaching the wall, bond orientational order is 
strongly suppressed and in fact we do not find any nucleation events happening in close proximity to the walls.
As a first approximation, density can be used as a measure of positional order, but more rigorous definitions
are also possible~\cite{russo_hs,mathieu_russo_tanaka}.
The range of the perturbation induced by the rough wall is governed by the correlation length of bond orientational order 
in the bulk phase.
It is well known that such structural correlation
lengths increase as the density is increased, but its absolute value is always very small (no static correlation length has been found that
exceeds a few particles diameters~\cite{tanaka,tanaka_review,kob2012non,mathieu_icosahedra,mathieu_russo_tanaka}), thus the value of 
the crossover is rather insensitive of the state point considered. We can thus conclude that the perturbation induced by the walls in our
system extends roughly only for distances up to $z\lesssim 3\sigma$, and what we observe are genuine homogeneous nucleation events.

\subsection{Gravity effects on crystal growth}\label{sec:growth}
In this section we address the question of how nuclei grow in the presence of a gravitational field. We already observed in
previous sections that the average position of the centers of mass of nuclei shifts to higher $z$ as the nuclei grow. We also argued that
this is due to the differences in the dynamics of the fluid particles on the two sides of the growing nuclei. The side
with higher $z$ is characterized by a faster dynamics and consequently a faster crystal growth.

\begin{figure}[!t]
 \centering
 \includegraphics[width=8cm,clip]{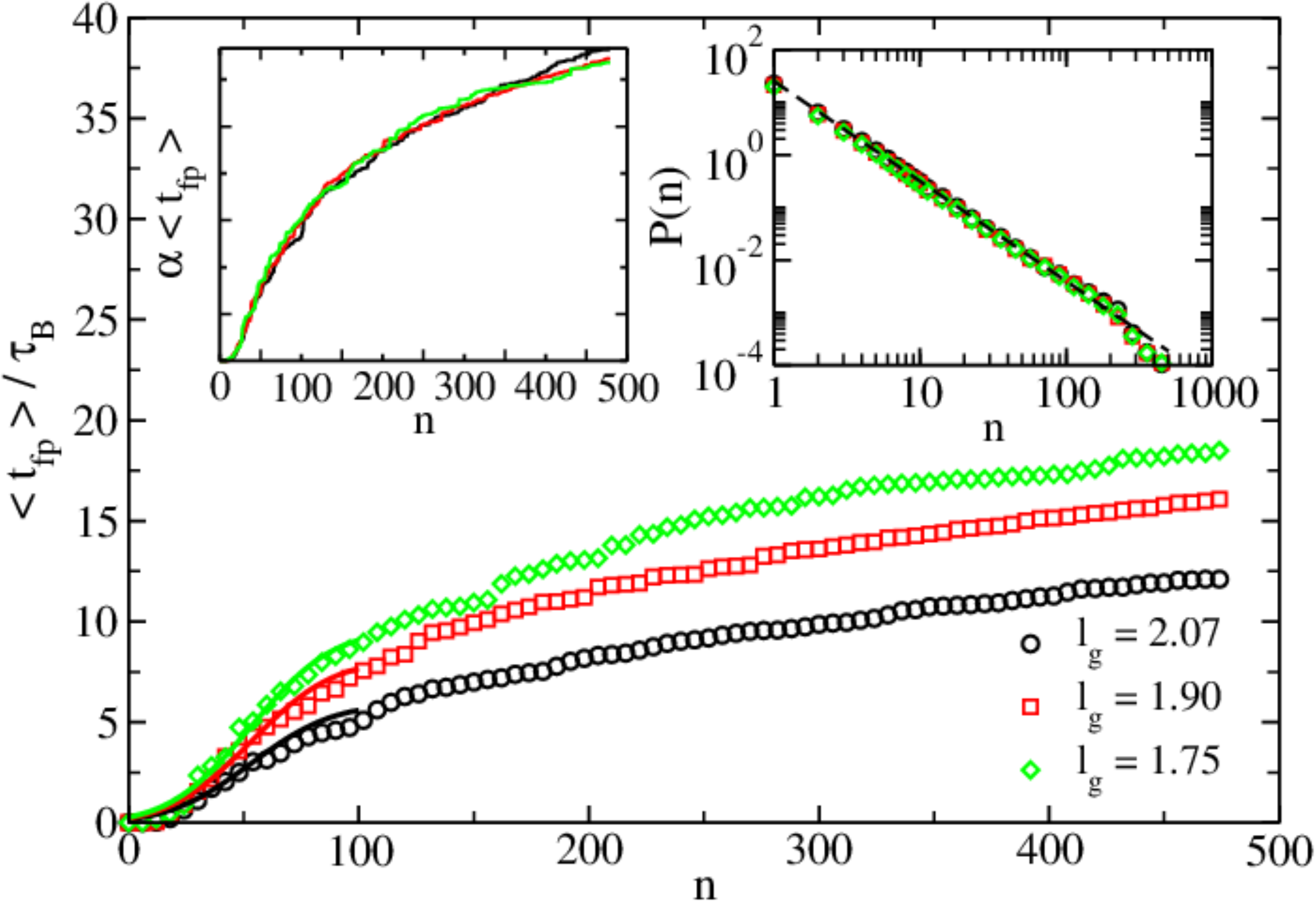}
 \caption{Mean first passage time as a function of the nucleus size $n$ for state points of group I. Symbols are measured
mean first passage times, whereas continuous lines are fits to Eq.~(\ref{eq:tfp}) up to $n=120$. 
The left inset shows that by scaling the times
all curve at different field strengths collapse on the same curve. The right inset shows the average distribution of crystal sizes $P(n)$
for all configurations in which the biggest cluster has size smaller than $400$ particles. The dashed line represents a power-law crystal size distribution
with Fisher exponent, $\tau=1.9$.}
\label{fig:mfpt}
\end{figure}

Figure~\ref{fig:mfpt} plots the mean first passage time for simulations of group I. The mean first passage time $\langle t_{fp}(n)\rangle$ is
defined as the average time elapsed until the appearance of a nucleus of size $n$ in the system. For homogeneous systems
it was shown that the following expression applies~\cite{wedekind_freeenergy}
\begin{equation}
 \langle t_{fp}(n)\rangle=\frac{1}{2kV}\left\{1+\text{erf}\left[c(n-n_c)\right]\right\}, \label{eq:tfp}
\end{equation}
where $k$ is the nucleation rate, $n_c$ is the critical nucleus size, $\text{erf}$ is the error function, 
and $c=\sqrt{\Delta F''(n_c)/k_{\rm B}T}$. 
Here $\Delta F''(n_c)$ is the second derivative of the nucleation barrier $\Delta F(n)$ at its maximum 
and thus $c$ characterizes the curvature at the top of the nucleation barrier profile. 
A direct fit is attempted for $n<120$ and represented as continuous lines in Fig.~\ref{fig:mfpt}. We limit the fit to small nuclei, since
as the nucleus grows it likely feels the effects of the density gradient, which Eq.~(\ref{eq:tfp}) does not take into account. 
Moreover the growth of the nucleus is also affected by the presence of surrounding smaller nuclei, as described in Ref.~\cite{valeriani2012compact}.
The fit gives us
nucleation times in good agreement with the one reported in Table~\ref{tab:state_points} and a critical nucleus size of approximately $50$
particles for all state points. The coincidence of the critical nucleus size is not surprising, as we have shown that nucleation
occurs for all state points in regions with similar volume fractions. Note the relation between the mean first passage time and the gravitational
length: longer gravitational lengths correspond to shorter mean first passage times, i.e. faster growth. This relationship is
much deeper: in the right inset of Fig.~\ref{fig:mfpt} we show that it is possible to collapse all mean first passage times by just rescaling
the time unit with a scaling factor $\alpha$. This rescaling can be explained in the context of mean first passage theory of activated processes, as developed in Ref.~\cite{wedekind_freeenergy} (we follow here its notation).
One first introduces the auxiliary function
$$
B(n)=\frac{1}{P_\text{st}(n)}\left[\int_n^b P_\text{st}(n')\,dn'-\frac{\langle t_{fp}(b)\rangle-\langle t_{fp}(n)\rangle}{\langle t_{fp}(b)\rangle}\right], 
$$
where $P_\text{st}(n)$ is the stationary time-independent probability of finding a nucleus of size $n$ and $b$ is the size at which simulations are stopped ($b=480$
in our case). First we note that $P_\text{st}(n)$ is the same for all state points reported in Fig.~\ref{fig:mfpt} (group I), since $P_\text{st}(n)$ depends
on the density accessible to the system, and not on the density gradient.
This is seen in the right inset of Fig.~\ref{fig:mfpt} which shows that even the full crystal size distribution $P(n)$, which includes both stationary
and non-stationary states with clusters bigger than the critical size, is unchanged for all state points. The decay of $P(n)$ is slow, and for the limited
sizes available to our study, it resembles a power law with Fisher exponent $\tau\simeq 1.9$. As we will see soon, the growth of the nucleus occurs faster laterally,
and this exponent can suggest a similarity with a two-dimensional percolation process (where $\tau=187/91$), in which the largest nucleus grows by merging with smaller nuclei.
A consequence of the observed scaling of $\langle t_{fp}(n)\rangle$ is that
all state points of group I are characterized by the same function $B(n)$.
Once the function $B(n)$ is known, one can reconstruct the free energy landscape from the expression~\cite{wedekind_freeenergy}
$$
\beta\Delta F(n)=\log{B(n)}-\int\frac{dn'}{B(n')}+C. 
$$
This means that the simulations with different gradients share the same free energy landscape, as already noted with the equivalence of the critical
nucleus sizes. $B(n)$ enters also into the definition of the generalized diffusion coefficient $D(n)$, which expresses 
the rate of attachment of particles to a nucleus of size $n$~\cite{wedekind_freeenergy}: 
$$
D(n)=B(n)/\frac{\partial\langle t_{fp}\rangle}{\partial n}. \label{eq:dn}
$$

The above theory provides a basis for understanding the effects of density gradient on the initial processes of crystallization shown in Fig. \ref{fig:mfpt}. 
Since we have established that $B(n)$ is the same for all simulations of group I, we conclude that the growth of the nuclei, as expressed by
the mean first passage time, is simply inversely proportional to the generalized diffusion $D(n)$.
We observed in Fig.~\ref{fig:mfpt} that shorter gravitational lengths are accompanied by slower growth, 
which is a consequence of smaller $D(n)$. 
Physically this corresponds to a more difficult growth of interfaces when
the density gradient is stronger. 
On noting that $D(n)$ is the rate of attachment of particles to a nucleus of size $n$, 
we speculate there are two origins behind the gradient-induced slowing down of crystal growth: a dynamical and a thermodynamic origin. 
First we consider the dynamical origin. The diffusion constant is a very strong decreasing function of $\phi$ near $\phi_g$.  
Thus the density gradient has a nonlinearly amplified strong perturbation on the dynamics. This should lead to a significant slowing 
down of particle diffusion on the high density side of a nucleus. On the other hand, the thermodynamic origin may play an important role in 
the slowing down of the growth on 
the opposite side (the low density side). The thermodynamic driving force decreases dramatically when the liquid density decreases 
toward the melting volume fraction, where the crystal and the liquid have the same free energy. Thus, we expect that the lower density 
of the liquid surrounding a crystal leads to the weaker driving force for crystal growth and thus to the slower growth.

\begin{figure}[!t]
 \centering
 \includegraphics[width=8cm,clip]{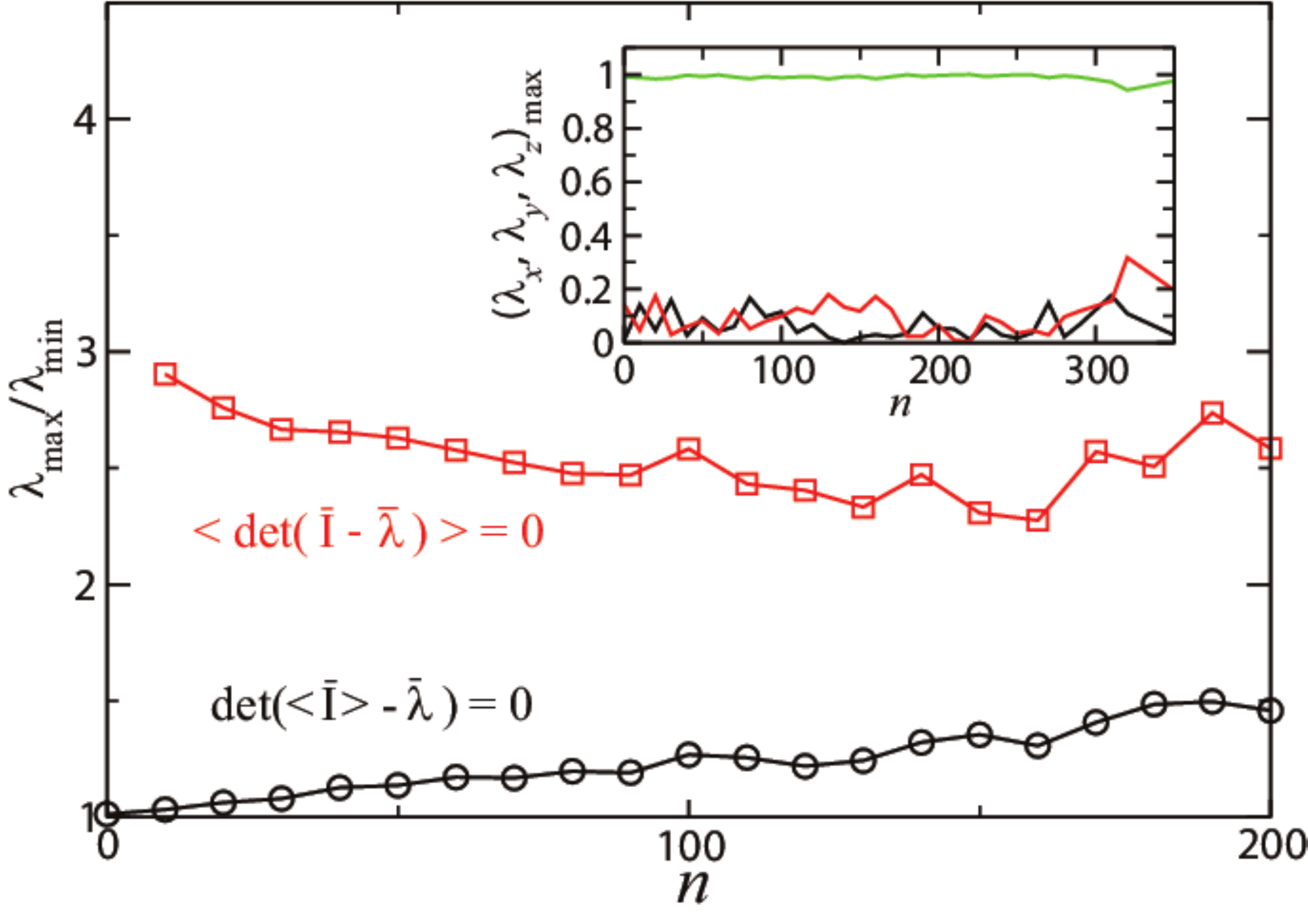}
 \caption{Shape of nuclei as a function of size $n$ for the state point of group II and $\phi_\text{avg}=0.520$. The shape is expressed as the
ratio between the maximum and minimum eigenvalue of the inertia tensor matrix. Two different types of averages are considered. The first
one is just the simple average over the eigenvalues of individual nuclei, and is represented by the square (red) symbols. With the round (black)
symbols we represent instead the ratio between the maximum and minimum eigenvalue of the averaged inertia tensor. The inset shows the $x$ (black),
$y$ (red) and $z$ (green) components of the eigenvector corresponding to the maximum eigenvalue of the averaged inertia tensor.}
 \label{fig:asphericity}
\end{figure}

We conclude this section by looking at the effects of the gravitational field on the shape and the orientation of the growing nuclei.
The shape can be determined by calculating the inertia tensor of nuclei
\begin{equation}
 I_{lm}=\sum_{i=1}^n |\vec{r}_i|^2\delta_{lm}-r_{i,l}r_{i,m},
\end{equation}
where $\vec{r}_i$ is the vector from particle $i$ to the center of mass of a nucleus, 
$l$ and $m$ are its vector components, and $\delta$ is the Kronecker delta.
The eigenvalues of the inertia tensor represent the inertia moments along the principal axis of inertia, given by the corresponding eigenvectors.
The ratio between the maximum eigenvalue and the minimum eigenvalue describes the asphericity of the crystalline nucleus.
In Fig.~\ref{fig:asphericity} we report the values of this ratio as a function of the size of the nuclei for two different types of averages.
The first one is simply the average of the ratio $\lambda_\text{max}/\lambda_\text{min}$ for individual nuclei, and is reported in the
square (red) symbols. It clearly shows that individual nuclei are always very aspherical. 
This comes not as a surprise, since the volume fraction 
at which the nuclei are forming is always rather high, and deviations from the spherical shape have already been reported at these
volume fractions~\cite{sanz,valeriani2012compact}. We observe that despite the aspherical shape, nuclei are always clearly distinct from each other:
we are still far from a spinodal type of nucleation. The second type of average is reported with the round (black) circles in
Fig.~\ref{fig:asphericity}, and it is the ratio $\lambda_\text{max}/\lambda_\text{min}$ for the average inertia tensor. Averaging
the inertia tensor of different nuclei corresponds to looking at the convolution of their shapes. If nuclei are aspherical but
randomly oriented, their convoluted shape will still spherical. This is exactly what we observe for small nuclei in Fig.~\ref{fig:asphericity},
where the ratio $\lambda_\text{max}/\lambda_\text{min}\sim 1$ for small $n$. As the nuclei grow the ratio increases steadily, and this is
due to an asymmetric growth induced by gravity. In the inset of Fig.~\ref{fig:asphericity} we report the components of
the principal axis of inertia (corresponding to the maximum eigenvalue) of the convoluted shape. Clearly this inertia axis is oriented
along the $z$ direction, i.e., along the gravity field. This means that the nuclei grow as ellipsoids with the two major axis laying in
the $(x,y)$ plane. A process which contributes to this result is probably also the merging of different nuclei in the $x,y$ directions,
as we have already shown that many nuclei form in a rather narrow $z$ strip.

\begin{figure}
 \centering
 \includegraphics[width=8.5cm,clip]{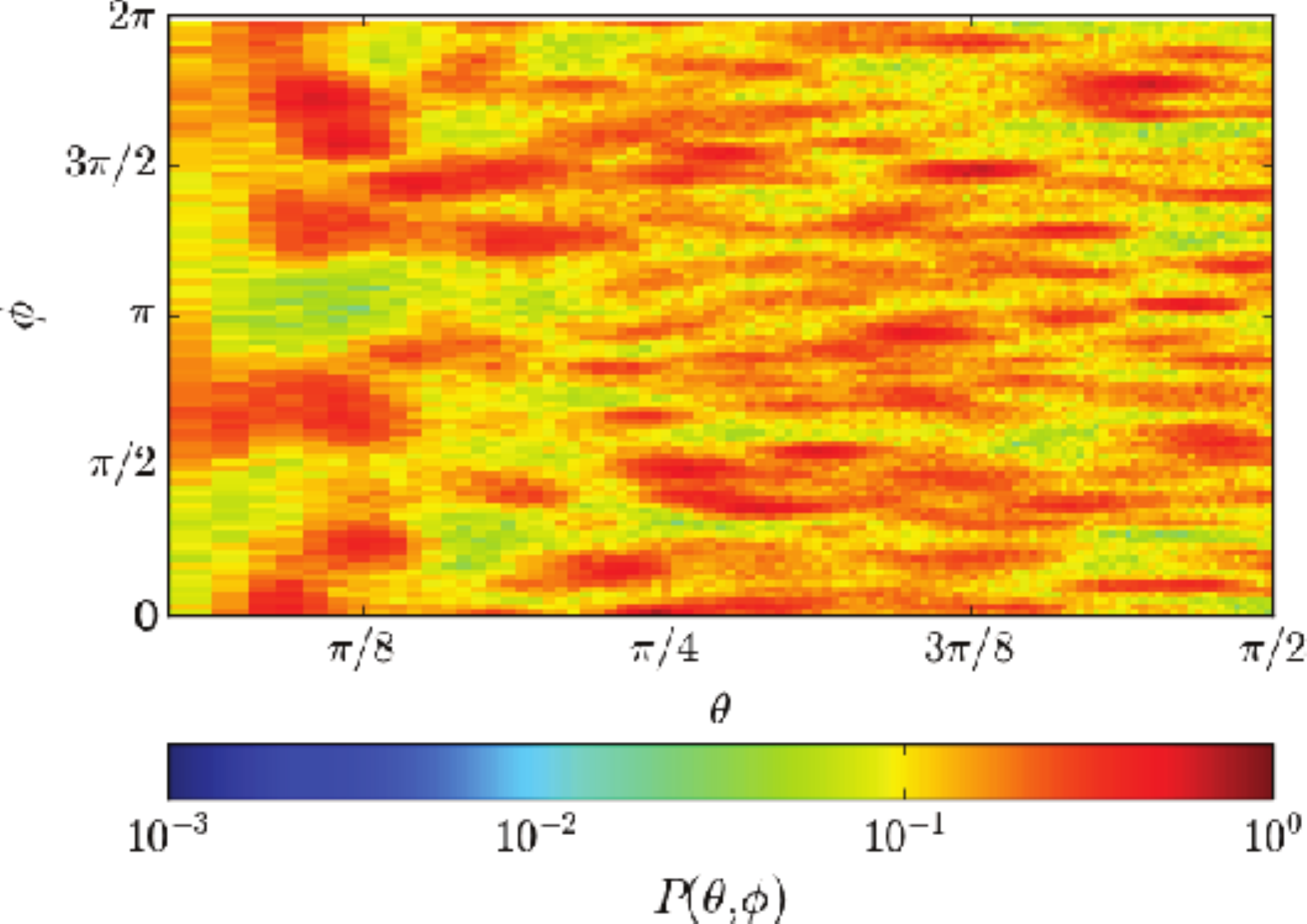}
 \caption{Probability distribution, in spherical coordinates $(\theta,\psi)$, for the orientation of the hexagonal plane of crystals
 formed at the state point with $l_G=1.75$ and $\phi_{\text{avg}}=0.520$.
 $\theta$ is the angle between the vector perpendicular to the plane and the $z$-axis (along which gravity is directed). $\psi$ is the angle between
 the projection on the $(x,y)$ plane of the vector perpendicular to the plane and the $x$ axis. Note that we consider weighted averages, where
 each nucleus enters in the definition of $P(\theta,\psi)$ with a weight equal to its size (similar results are obtained with unweighted averages).}
 \label{fig:hexagonal_map}
\end{figure}
In Fig.~\ref{fig:hexagonal_map} we consider the orientation of crystal planes for the state point with $l_G=1.75$ and $\phi_{\text{avg}}=0.520$.
It is well known that for hard potentials the relevant crystal polymorphs are either
$fcc$ or $hcp$ (and $rhcp$ which is given by randomly stacking $fcc$ and $hcp$ planes)~\cite{russo_hs}. Both polymorphs are characterized by
hexagonal planes. For $fcc$ the hexagonal plane is written as $(1,1,1)$ in Miller indices (due to the $C_4$ symmetry of cubic crystals,
there are actually $4$ planes differing for a $\pi/2$ rotation along any of the unit cell vectors). For $hcp$ the hexagonal plane
is written as $(0,0,0,1)$ in Miller-Bravais indices. For each crystalline particle in a nucleus we detect the direction of the hexagonal
plane (the vector perpendicular to the plane) and plot its probability distribution in spherical coordinates, according to the usual
transformations: $r=\sqrt{x^2+y^2+z^2}$, $\theta=\cos^{-1}(z/r)$ and $\psi=\tan^{-1}(y/x)$ ($z$ is the direction of gravity).
The probability to find a crystalline particle with hexagonal planes pointing in the $(\theta+d\theta,\psi+d\psi)$ direction is then given by
$P(\theta,\psi)\sin\theta\,d\theta\,d\psi$.
We have approximately $50$ independent trajectories, and for each we analyse the orientation of crystal particles belonging only to the largest cluster in
the system, and only if the cluster has size bigger than $20$ particles (to avoid the contribution from metastable nuclei).
In Fig.~\ref{fig:hexagonal_map} the peaks corresponding to the orientation of the individual crystals are still visible, but it is already clear that
$P(\theta,\psi)$ has no sensible $\psi$ dependence.

\begin{figure}
 \centering
 \includegraphics[width=8cm]{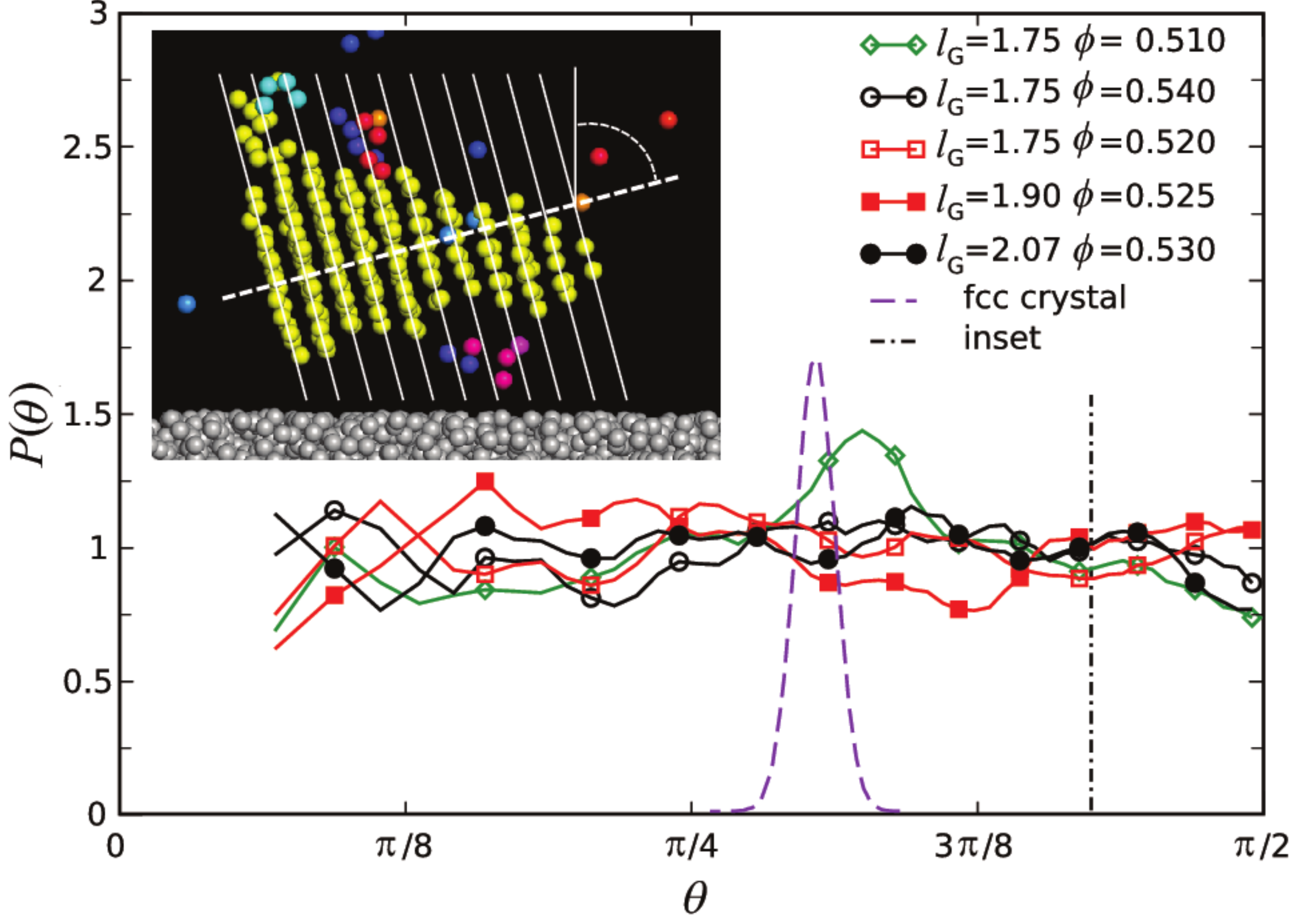}
 \caption{Probability distribution function $P(\theta)$ for all state points of group I and II (continuous lines with symbols), scaled so
 that $P(\theta)=1$ represents a uniform distribution.
 The dashed line shows the probability distribution for a bulk $fcc$ crystal with lattice vectors oriented along the $(x,y,z)$ directions and at volume fraction $\phi=0.535$
 (the distribution function is scaled to improve readability).
 The inset shows a snapshot from a nucleation event at $l_G=1.75$ and $\phi_{\text{avg}}=0.520$: the continuous lines are traced along the hexagonal
 planes while the dashed line gives the plane orientation. The $\theta$ angle of the nucleus in the inset is shown as the dashed-dotted line in the main panel.}
 \label{fig:hexagonal_theta}
\end{figure}
To examine the $\theta$ dependence, in Fig.~\ref{fig:hexagonal_theta} we plot the reduced probability distribution $P(\theta)=\int P(\theta,\psi)\,d\psi$.
This plot shows that, for all state points in groups I and II,
there is no noticeable $\theta$ dependence for the orientation of the crystalline planes. This means that the nucleation stage occurs homogeneously,
with nuclei having no preferred orientation. Since the rotational diffusion of nuclei is much slower than the growth process,
the nuclei retain their random orientation even when the average shape of the nuclei becomes asymmetric (Fig.~\ref{fig:asphericity}).
One example of crystal orientation and of its inclination $\theta$ is shown in the inset of Fig.~\ref{fig:hexagonal_theta} (the same value
of $\theta$ is indicated in the main panel as a dashed-dotted line).
In Fig.~\ref{fig:hexagonal_theta} we note that for the state point where the average nucleation event is closest to the wall
($l_G=1.75$ and $\phi_{\text{avg}}=0.510$), there is a small probability excess close to $\theta=\cos^{-1}(1/\sqrt{3})$.
This orientation corresponds to a cubic crystal oriented with its lattice vectors along the $(x,y,z)$ directions, as shown in the dashed curve
for a thermal $fcc$ crystal. We can speculate that, for nucleation events occurring very close to the wall (low values of $\phi_\text{avg}$ and
high $G$ values) the orientation of crystals could become anisotropic, but a confirmation of this effect needs more statistical significance.

\subsection{Comparison with experiments}\label{sec:experiments}

\begin{table*}[!t]
  \small
  \caption{Comparison of colloidal diameter $d$, colloidal type and density $\rho_P$, solvent type and density $\rho_f$,
    and gravitational lengths $l_G$ for the experiments in Refs.~\cite{schatzel1993density,harland1997crystallization,sinn2001solidification,iacopini,franke2011heterogeneous}.
    It should be noted that the determination of gravitational lengths in experiments is subject to high uncertainty. The determination of
    the size of particles is especially difficult, and some estimates indicate that the error is of the order of $3-6\%$~\cite{royall2013search}.}
  \label{tab:experiments}
  \begin{tabular*}{0.8\textwidth}{@{\extracolsep{\fill}}lllll}
    \hline
    \emph{experiment} & \emph{colloids} & $d$ & \emph{solvent} & $l_G/d$  \\
    \hline
    Sch{\"a}tzel et al.~\cite{schatzel1993density} & PMMA & $1\,\mu m$ & decalin/tetralin& $2.9$ \\
    &  $\rho_P=1.19 g/cm^3$ &  & $\rho_f=0.92 g/cm^3$ & \\ \hline
    Harland \& van Megen~\cite{harland1997crystallization} & PMMA & $0.40\,\mu m$ & decalin/CS$_2$ & $138$ \\
    &  $\rho_P=1.19 g/cm^3$ &  & $\rho_f=0.97 g/cm^3$ & \\ \hline
    Sinn et al.~\cite{sinn2001solidification} & PMMA & $0.89\,\mu m$ & decalin/tetralin & $4.1$ \\
    &  $\rho_P=1.19 g/cm^3$ &  & $\rho_f=0.92 g/cm^3$ & \\ \hline
    Iacopini et al.~\cite{iacopini} & polystyrene microgel & $0.86\,\mu m$ & 2-ethyl-naphthalene & $80$ \\
    Franke et al.~\cite{franke2011heterogeneous} &  $\rho_P=1.01 g/cm^3$ &  & $\rho_f=0.992 g/cm^3$ & \\ \hline
  \end{tabular*}
\end{table*}

\begin{figure}[!t]
  \centering
  \includegraphics[width=8cm,clip]{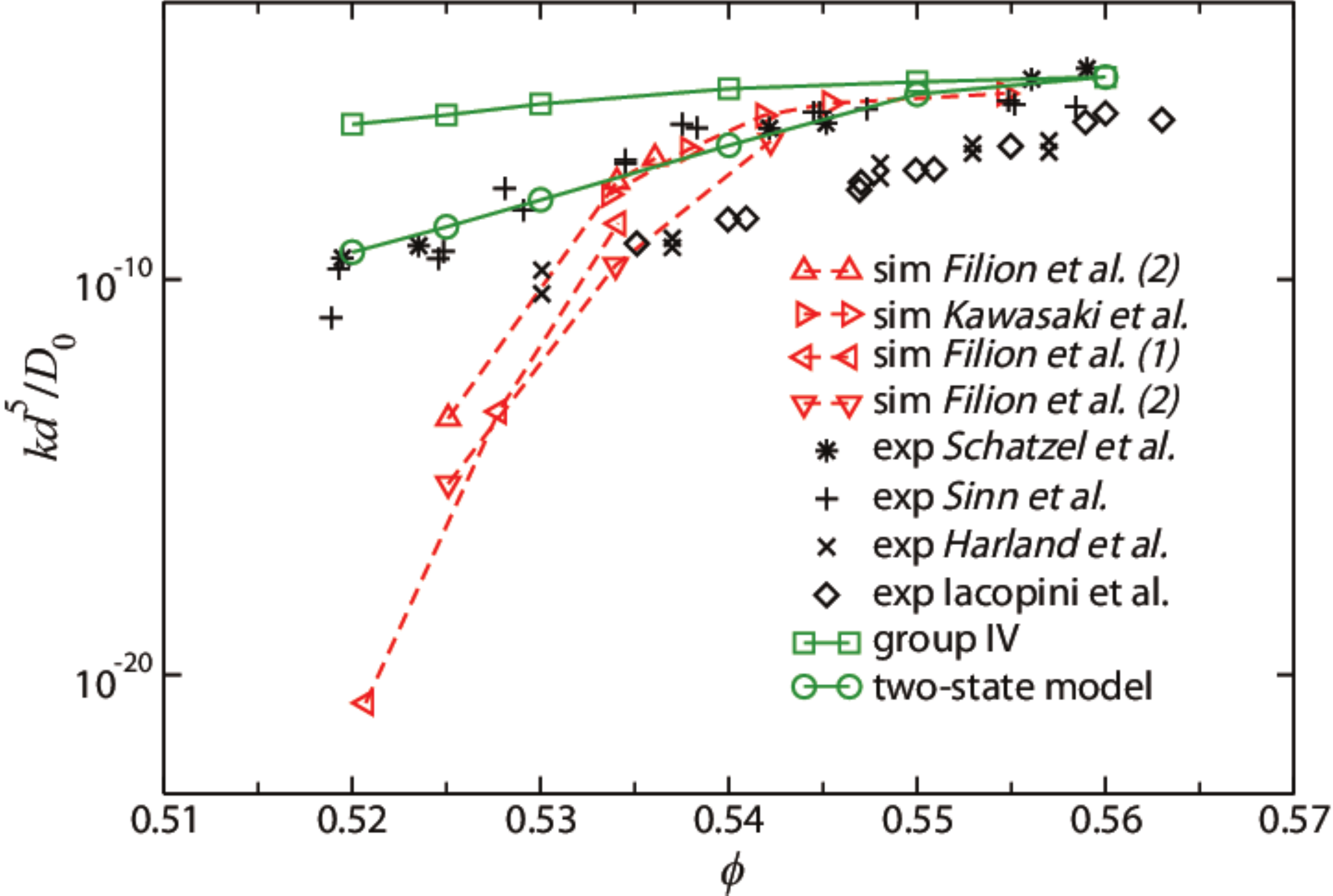}
  \caption{Adimensional crystal nucleation rates estimated from
    simulations (dashed red lines) and from experiments
    (black symbols). The legends have the following correspondence:
    Filion et al.(1) is Ref.~\cite{filion}, Filion et al.(2) is
    Ref.~\cite{filion2}, Kawasaki et al. is Ref.~\cite{kawasaki},
    Schatzel et al. is Ref.~\cite{schatzel1993density} ($l_G=2.9\,d$), Harland et
    al. is Ref.~\cite{harland1997crystallization} ($l_G=138\,d$), Sinn et al. is
    Ref~\cite{sinn2001solidification} ($l_G=4.1\,d$) and Iacopini et al. is Ref.~\cite{iacopini} ($l_G=80\,d$).  Nucleation rates for the
    simulations of group IV and the two-state model fit are
    reported as continuous green lines.  This figure is drawn starting
    from Fig.~6 of Ref.~\cite{filion2} and Fig.~12 of
    Ref.~\cite{harland1997crystallization}.  }
  \label{fig:nucleationrate}
\end{figure}

We now address the question whether a gravitational field can enhance the crystallization rate in a colloidal suspension of
hard spheres. We first report in Table~\ref{tab:experiments}
some experimental parameters relevant to our study. The experiments can be clearly distinguished according
to their gravitational lengths $l_G$. Experiments in Refs.~\cite{schatzel1993density,sinn2001solidification} involve
colloidal particles suspended in an index-matched solvent but not in a density matched one, resulting in very short
gravitational lengths. Experiments in Refs.~\cite{harland1997crystallization,iacopini,franke2011heterogeneous} instead improve
considerably the density matching by either employing small particles, or by using swelling microgels whose density is very close
to the density of the solvent. In Fig.~\ref{fig:nucleationrate} we compare the adimensional nucleation rates as a function of
volume fraction calculated in these experiments (all experimental results are plotted with black symbols).
Experiments with shorter gravitational lengths (plus symbols~\cite{sinn2001solidification} and stars~\cite{schatzel1993density})
are characterized by higher nucleation rates when compared to experiments with longer gravitational lengths
(crosses~\cite{harland1997crystallization} and diamonds~\cite{iacopini}). This shows that a reduction of the
gravitational effects goes indeed in the right direction of explaining the discrepancy between experiments and simulations.

To assess the importance of gravitational effects in the crystallization of hard spheres
we define the following quantity:
$Q(l)=\tau_s/\tau_x$, where $\tau_s$ is the average time for a colloid
to move the distance $l$ due to the gravity field, and $\tau_x$ is the
average time for a nucleation event to occur in the volume $l^3$.
$Q(l)$ is thus an adimensional number which quantifies the relative
importance of the sedimentation timescale with respect to the
crystallization timescale, at any particular length scale $l$.  
For $Q(l)\gg 1$ we expect gravitational effects to be
negligible, and the observed nucleation rate in experiments to be the
same as in gravity-free simulations. On the other hand, for $Q(l)\ll
1$ gravitational effects cannot be ignored as they become significant
on timescales much shorter than the average nucleation time.  The most
relevant length scale $l$ in this problem is the size of the critical
nucleus $R_c$, since below that size, $l<R_c$, nuclei can convert back
into the metastable melt. We will thus focus on $Q(R_c)$ at ordinary
experimental conditions.  Table~\ref{tab:experiments}
reports some experimental parameters relevant to the determination of
$Q(R_c)$, namely, the diameter of the colloids $d$, the density of the
solvent $\rho_f$, and the gravitational length $l_G$, and the density of a colloidal particle $\rho_P$.
The details of the calculations are given in the Appendix~\ref{appendix:A}. We find that the condition
$Q(R_c)\sim 1$ is realized in a small windows of $|\Delta\mu|$ (the chemical potential
difference between the solid and fluid phase),
$\beta |\Delta\mu|\sim 0.38$, and $\phi\sim 0.525$, for the experiments in
Refs.~\cite{schatzel1993density,sinn2001solidification}.  These values
are slightly less (as expected) for the experiments with a longer
gravitational length, Refs.~\cite{harland1997crystallization,iacopini},
i.e. $\beta |\Delta\mu|\sim 0.36$ and $\phi\sim 0.522$. Thus, for
$\phi\gtrsim 0.525$, we find $Q(R_c)\gg 1$ and gravitational effects
can be ignored. But for $\phi\lesssim 0.525$ the converse is true, and
gravitational effects become increasingly important.
Thus, for $\phi$ significantly larger
than $0.525$
we expect
that experiments without density matching (Refs.~\cite{schatzel1993density,sinn2001solidification}) and gravity-free simulations will measure similar
nucleation rates, whereas a big discrepancy,
due to gravitational effects, should emerge at $\phi\lesssim 0.525$.
This can be confirmed by looking at the nucleation rates in Fig.~\ref{fig:nucleationrate}
where experiments are plotted with (black) symbols, while simulations
without gravity as (red) lines and symbols, confirming that
the value obtained from our simple dimensional analysis, $\phi\sim
0.525$, is indeed between these two regimes.

\begin{figure}[!t]
  \centering
  \includegraphics[width=8cm,clip]{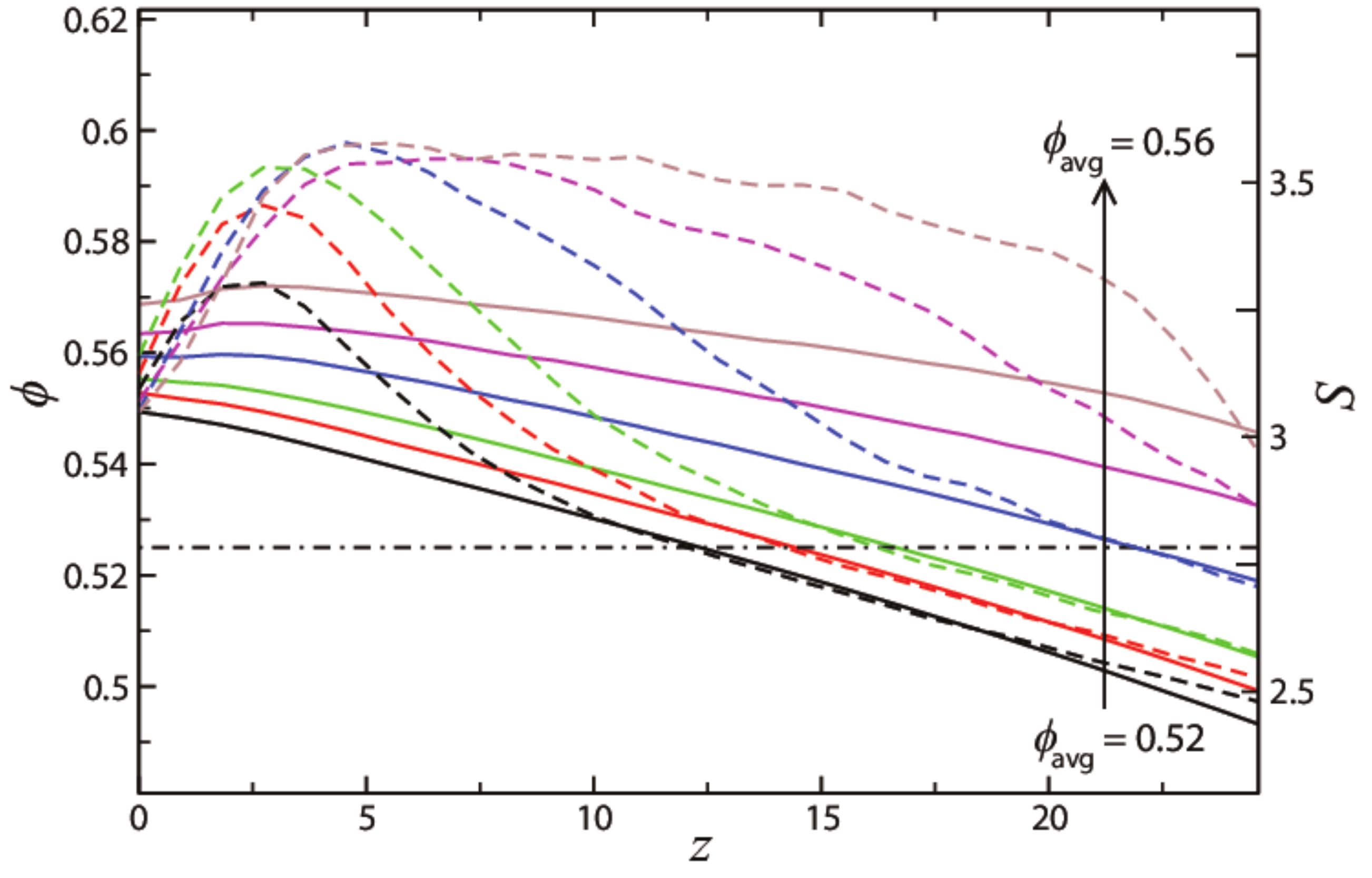}
  \caption{Volume fraction profiles $\phi(z)$ (full lines) and
    crystallinity profiles $S(z)$ (dashed lines) for state points in
    Group IV.  The $\phi$ scale and the $S$ scale
    are reported respectively on the left and right axis.  The nearly
    horizontal dashed-dotted line marks the value $\phi=0.525$,
    separating the region of crystal formation from the metastable
    region.  Profiles are obtained by averaging all configurations in
    which the biggest nucleus size is between $20$ and $30$ particles,
    and by dividing the $z$ dimension into bins of size $\Delta z=1$.}
  \label{fig:phi_profile2}
\end{figure}
The same behaviour is seen within our simulations.
In Fig.~\ref{fig:phi_profile2} we plot the $z$ profiles of both volume fraction,
$\phi$, and crystallinity, $S$, for the state points in group IV.
The profiles are taken by averaging
all configurations in which the largest nucleus has a size comprised
between $20$ and $30$ particles, in order to have a picture of the
nucleation process in its early stage. The figure reveals that, at the
beginning of the nucleation events, a $z$-dependent profile has
developed both for $\phi$ and $S$. While $\phi$ has a smooth monotonic
behavior, apparently unaffected by the ongoing crystallization
process, the crystallinity order parameter $S$ reveals that
the location of the nucleation events is in the density enhanced
regions. The extent of these regions depends on the average volume
fraction, $\phi_\text{avg}$. This is shown by the dashed-dotted line
in Fig.~\ref{fig:phi_profile2} which clearly separates two regimes: for
$\phi\lesssim 0.525$ the $\phi$ and $S$ profiles display the same $z$
dependence, while $\phi\gtrsim 0.525$ marks the beginning of the
nucleation events. We recall from our previous adimensional analysis
that $Q(\phi= 0.525)\sim 1$, again confirming that for $\phi\gtrsim
0.525$ nucleation events are bulk-like and the same as in a
gravity-free environment, whereas for $\phi\lesssim 0.525$
sedimentation can occur on shorter time-scales than nucleation, and
significant deviations are to be expected with respect to the zero
gravity case.
The nucleation process under gravity is inevitably out of equilibrium and even
hydrodynamics should play an important role eventually.  However, we
argue that within the incubation time (at most $\sim 10^2$ Brownian
times) there may be no macroscopic processes involved and
gravity-induced density fluctuations via diffusion may be a major
process.
This is indirectly shown by the experiments in Ref.~\cite{schatzel1993density}
which report that the first indication of crystallization could be
observed on timescale of $10^3$ s with a solvent viscosity of
$2.37\cdot 10^{-3}$ Pa$\cdot$s. This corresponds to incubation times
of the order of $10^2$ Brownian times, which is the same range
measured in our simulations.
Despite having similar incubation times, experiments in non-density matched
solvents and simulations differ for their nucleation rates, as can be seen
in Fig.~\ref{fig:nucleationrate}, where the nucleation rates of simulations
in group IV are reported as (green) squares. But this difference is trivially
due to the different volumes accessible in simulations and experiments (the nucleation
rate is obtained by dividing the average incubation time by the total volume of the system).
Simulations measure nucleation events in strips of height $z$, while
experiments measure nucleation events in regions of height $\sim
10^4\,z$ (the section of the laser beam), so that the difference in
nucleation rates between experiments and simulations at the lowest volume
fraction is expected to be of the order of $10^4$, provided that the
experiments are sensitive enough to detect the formation of only a
few nuclei.  This estimation well matches with the ratio in the
nucleation rate between the experiment and our simulation observed
at $\phi=0.52$, as shown in Fig.~\ref{fig:nucleationrate}.  The
physical picture which emerges is thus that, at low volume fraction,
the nucleation rate is controlled by small density inhomogeneities
induced by gravity. On small scales these inhomogeneities should
resemble the ones obtained in simulations.

Given the previous physical picture, we can easily build a model
to connect the results at high $\phi$ (where bulk crystallization
dominates) and low $\phi$ (where sedimentation dominates).
We adopt a simple two-state model, with high-density regions
($\phi>\phi^*$ and with nucleation rates similar to the ones
extracted from our simulations) coexisting with low-density regions
($\phi<\phi^*$, and with nucleation rates similar to the bulk
behavior in absence of gravity).  Due to the steep increase in
nucleation rates we can take the value $\phi^*$ as the density of
the nucleation rate maximum, $\phi^*\sim 0.56$.
The nucleation rate in the sample can thus be written as
$k=k_Sx+k_H(1-x)$, where $k_S$ is the rate extracted from our
simulations, $k_H$ is the rate obtained without gravity, and $x$ is
the fraction of the volume in the sample with $\phi>\phi^*$ due to
gravity (and not thermal fluctuations).  We model the $\phi$
dependence of $x$ by a Fermi function to account for the constraint
on $x$ from the conservation of the total volume fraction $\phi$:
$x(\phi)=1/(1+\exp\{\kappa(\phi-\phi^*)/G\})$, so that at $G=0$
density inhomogeneities are null, while for $G>0$ the extent of the
fluctuations is proportional to $\exp{(\phi-\phi^*)}$. $\kappa\sim
1.5$ is fixed from the equivalence of nucleation rates at
$\phi=0.52$, as previously discussed.  The results of the model are
depicted in Fig.~\ref{fig:nucleationrate} as a dashed line for the
experiments in
Refs.~\cite{schatzel1993density,sinn2001solidification}.  As
expected, for $\phi>0.525$ the nucleation rate gradually recovers
its gravity free value with an increase in $\phi$.  The good
agreement shows that, at least in principle, nucleation enhanced by
gravity-induced density fluctuations is a viable mechanism to
explain the discrepancy between experimental and theoretical
results.

\section{Conclusion}\label{sec:conclusions}
In the previous sections we have considered the interplay between sedimentation and
crystallization in a model of colloidal hard spheres. Gravity is a very important
factor that determines the crystallization behaviour in many experimental situations~\cite{zhu1997crystallization}:
as we have shown, even density-matched suspensions are characterized by rather small
gravitational lengths (see Table~\ref{tab:experiments}).

The first noticeable effect of gravity is the
strong enhancement of nucleation rates, which is due to the increase of the local density
in proximity of the walls. Nucleation events occur preferentially in regions where, due to sedimentation,
the volume fraction is approximately $55-56\%$, in correspondence of the nucleation
rate maximum in bulk hard spheres. In this respect, the nucleation process is similar to a
homogeneous nucleation event, with similar nucleation rates, and with pre-critical nuclei which are on
average spherical and have crystal planes randomly oriented with respect to the direction of gravity.
The symmetry breaking induced by the gravitational field is seen in the growth stage, 
where a steeper density profile (shorter gravitational length) slows down the dynamics of the growth process, 
as seen by the reduction of the generalized diffusion coefficient $D(n)$.
The bottom side of the nucleus is in contact with a slowly relaxing fluid, while on the opposite side
the dynamics is much faster, leading to an increase of the average height of the center of mass
position as the nuclei grow. But the faster dynamics on the top side of the nucleus is eventually
compensated by a smaller thermodynamic driving force to crystallization, due to the decrease of density
along the $z$ direction. On average thus the nuclei will grow faster laterally, as shown by the
study of the average inertia tensor. As the nuclei grow, they become on average more asymmetric, with
their principal axis of inertia located along the $z$ axis, which again signals a faster growth on the
$x,y$ plane. An important contribution to crystal growth is also the merging of different nuclei
along the $x,y$ plane, as revealed by the distribution of crystal sizes.
The orientation of crystalline planes remains isotropic also in the growth stage, as the
rotational diffusion of nuclei is a slower process compared to their growth.

We devoted special attention to the study of the effects of rough walls.
By predicting the density profile from the equation of state, we were able to prepare walls at thermodynamic conditions close to
the nearby fluid, thus minimizing the disturbance introduced by the walls on the liquid structure.
First we determined that the effects of the walls on the dynamic properties of the fluid vanish on a length scale
comparable to the static correlation length in the bulk fluid. Close to the walls the dynamics is
greatly slowed down, and a decoupling of lateral and perpendicular diffusion occurs. These dynamic anomalies are accompanied
by a suppression of bond orientational order. This is the structural origin of the suppression of crystallization close to the
walls, and confirms previous simulations where it was shown that nucleation is mainly controlled by the development of bond 
orientational order~\cite{russo_hs}. Positional order, i.e. density, is instead almost unaffected by the presence of the walls,
providing a clean example where slowness is linked to
many-body correlators (like bond-orientational order) and not to two-body quantities (like density)~\cite{mathieu_russo_tanaka}.

Finally we looked at the experimental results on the crystallization of hard sphere suspensions in the
light of the gravitational effects, which we believe do play a major role in non-density matched
samples. We first identified the regime where sedimentation is possibly controlling the crystallization behaviour,
and showed that density inhomogeneities induced by the gravitational field are indeed
capable of enhancing the nucleation rate up to the values reported in the literature.
However, there are other non-ideal features in experiments, such as the presence of effects of shear flow
or other hydrodynamic effects, which our simulations do not take into account.
In order to single out unambiguously the mechanism responsible for the discrepancy between simulations and experiments, 
experiments with improved density matching should be carried out, possibly
showing a significant decrease in the nucleation rates.
Already the results of some experiments~\cite{harland1997crystallization,iacopini,franke2011heterogeneous,jade_royall} suggest that this
might be a promising mechanism, and we hope that the present work will stimulate more efforts towards this direction.


\section*{Appendix: Calculation of $Q(l)$}\label{appendix:A}
For
hard spheres $Q(l)$ can be immediately calculated as follows. $\tau_s$
is given by the Richardson-Zaki
expression~\cite{richardson1954sedimentation} for hindered settling at
low Reynolds numbers: $\tau_s=l\Xi/G(1-\phi)^{4.65}$, where $\Xi$ is
the Stokes drag coefficient and $G$ is the gravitational pull on the
colloids. To obtain $\tau_x$ we need an estimate of the nucleation
rate $k$ in hard-spheres.  This can be calculated within the framework
of Classical Nucleation Theory (CNT), where the nucleation rate $k$ is
simply the product of a kinetic term $K$ and a thermodynamic term $U$,
the former expressing the mobility of the fluid-solid interface, and
the latter accounting for the free energy barrier of formation of a
crystal nucleus.  For the kinetic term we use the expression
$K=\rho_fZf_c^+$, where $\rho_f$ is the density of the suspending
fluid, $Z=\sqrt{\beta|\Delta\mu|/6\pi n_c}$ is the Zeldovich factor,
and $f_c^+$ is the attachment rate of particles to the critical
cluster containing $n_c$ particles, usually written as
$f_c^+=24Dn_c^{2/3}/\lambda$. In the previous expressions
$|\Delta\mu|$ is the chemical potential difference between the solid
and fluid phase, $D$ is the short-time diffusion coefficient,
$\lambda$ is the typical distance over which diffusing particles
attach to the interface (which we set as a fraction of the particle's
diameter $\lambda=0.4\,d$ as was determined in
Ref.~\cite{auer2004quantitative}), and $\beta=1/k_{\rm B}T$. The
thermodynamic term of the nucleation rate is simply given by the free
energy barrier of formation of the critical nucleus,
$U=\exp(-\beta\Delta G_c)$.  We model the free energy with the CNT
expression, corrected with a radius ($R$) dependent interfacial free
energy $\gamma(R)$, namely $\Delta G=4\pi
R^2\tilde{\gamma}(1-\tilde{\epsilon}/R^2)-4\pi
R^3\rho_s|\Delta\mu|/3$, where $\tilde{\gamma}$ and $\tilde{\epsilon}$
are model-dependent constants, and $\rho_s$ is the density of the
solid phase. In Ref.~\cite{pristipino_cnt}, the values
$\beta\tilde{\gamma}d^2=0.741$ and $\tilde{\epsilon}/d^2=-0.279$ were
shown to describe very accurately the hard-spheres case. Combining the
above expressions for $\tau_s$ and $\tau_x$ we obtain $Q(l)$ as a
function of $\phi$ and $|\Delta\mu|$, which can be further simplified
by using an equation of state $|\Delta\mu|(\phi)$, which we derived by
a fit to simulation results~\cite{filion} in the $\phi$-range of
interest.

\section*{Acknowledgements}
This work was partially supported by a Grant-in-Aid for Scientific Research (S) from JSPS,  
Aihara Project, the FIRST program from JSPS, initiated by CSTP, and a JSPS Postdoctoral Fellowship.

\footnotesize{
\providecommand*{\mcitethebibliography}{\thebibliography}
\csname @ifundefined\endcsname{endmcitethebibliography}
{\let\endmcitethebibliography\endthebibliography}{}

}

\end{document}